\pdfoutput=1
\documentclass[reprint,12pt,onecolumn,notitlepage,nofootinbib,floatfix]{revtex4-2}
\usepackage{amssymb}
\usepackage{amsmath}
\usepackage{amsfonts} 
\usepackage{bm}
\usepackage{color} 
\usepackage{subfigure}
\usepackage{graphicx}
\usepackage[bookmarks=false]{hyperref} 
\usepackage{physics}
\usepackage{tikz}
\usepackage{multirow}
\hypersetup{pdfstartview=FitH,pdfhighlight=/O,colorlinks=true}

\allowdisplaybreaks

\usepackage{epsfig} 
\usepackage{epstopdf}
\usepackage{latexsym}
\usepackage{psfrag}   
\usepackage{subfigure}  
\usepackage{booktabs}  
\usepackage{braket}
\usepackage{mathtools}
\usepackage{textcomp}
\usepackage{ifthen}
\usepackage{tensor}
\usepackage{protosem} 
\usepackage{wasysym}

\usepackage[toc]{appendix}
\usepackage{color,soul} 
\usepackage{datetime}

\definecolor{darkblue}{rgb}{0.2, 0, 0.8}
\definecolor{darkgreen}{rgb}{0.2, 0.71, 0}
\definecolor{awesome}{rgb}{1.0, 0.13, 0.32}
\definecolor{cadmiumred}{rgb}{0.89, 0.0, 0.13}
\definecolor{dukeblue}{rgb}{0.0, 0.0, 0.61}

\newcommand{\bea}{\begin{eqnarray}}
\newcommand{\eea}{\end{eqnarray}}
\newcommand{\ba}{\begin{eqnarray}}
\newcommand{\ea}{\end{eqnarray}}

\newcommand{\beq}{\begin{equation}}
\newcommand{\eeq}{\end{equation} }
\newcommand{\beqa}{\begin{eqnarray}}
\newcommand{\eeqa}{\end{eqnarray}}
\newcommand{\beqar}{\begin{eqnarray*}}
\newcommand{\eeqar}{\end{eqnarray*}}

\renewcommand{\href}[2]{#2}

\begin{document}

\title{RRR fluctuations from non-singular bounce }

\author{Sayantan Choudhury}
\email{sayanphysicsisi@gmail.com, sayantan\_ccsp@sgtuniversity.org (Corresponding Author)}
\affiliation{Centre For Cosmology and Science Popularization (CCSP),
        SGT University, Gurugram, Delhi- NCR, Haryana- 122505, India }

\date{\today}

\begin{abstract}  \vskip 0.2in 

We present a consistent treatment of regularization, renormalization, and resummation in a framework involving a contracting phase, a non-singular bounce, and inflation with a sharp transition from slow-roll to ultra-slow-roll, enabling primordial black hole production. Within an effective field theory approach, we compute quantum loop corrections to the power spectrum across all phases. Quadratic UV divergences are removed through renormalization, while logarithmic IR divergences are softened via resummation, demonstrating scheme independence. The inclusion of contraction and bounce allows the generation of solar-mass PBHs, $M_{\rm PBH}\sim{\cal O}(M_\odot)$, with minimal inflation $\Delta N_{\rm Total}\sim{\cal O}(60)$, thereby evading the no-go theorem. For $0.88\le c_s\le1$, the peak spectrum amplitude lies within $10^{-3}\le A\le10^{-2}$, preserving causality and unitarity. PBH abundances across $10^{-33}\!-\!10^{-27}$ and $10^{-6}\!-\!10^{-1}M_\odot$ are confronted with microlensing constraints to control PBH overproduction.

\vskip 0.2in
\centering
\noindent {\it \footnotesize {\bf Essay written and received honorable mention 
for the Gravity Research Foundation 2026 Awards for
Essays on Gravitation}}

\end{abstract}   

\maketitle 

\newpage

\section{Introduction}

The concepts to create primordial black holes (PBHs) in various early Universe physics contexts have grown significantly in recent years due to the interest in investigating PBHs \cite{Zeldovich:1967lct,Hawking:1974rv,Carr:1974nx,Carr:1975qj,Chapline:1975ojl,Carr:1993aq,Choudhury:2011jt,Yokoyama:1998pt,Kawasaki:1998vx,Rubin:2001yw,Khlopov:2002yi,Khlopov:2004sc,Saito:2008em,Khlopov:2008qy,Carr:2009jm,Lyth:2011kj,Drees:2011yz,Drees:2011hb,Ezquiaga:2017fvi,Kannike:2017bxn,Hertzberg:2017dkh,Pi:2017gih,Gao:2018pvq,Dalianis:2018frf,Cicoli:2018asa,Choudhury:2013woa,Sasaki:2016jop,Raidal:2017mfl,Papanikolaou:2020qtd,Ashoorioon:2022raz,Papanikolaou:2022chm,Papanikolaou:2023crz,Wang:2022nml,Riotto:2023hoz,Riotto:2023gpm,Papanikolaou:2022did,Choudhury:2023vuj, Choudhury:2023jlt, Choudhury:2023rks,Choudhury:2023hvf,Choudhury:2023kdb,Choudhury:2023hfm,Bhattacharya:2023ysp,Choudhury:2023fwk,Choudhury:2023fjs,Choudhury:2024one,Choudhury:2024ybk,Choudhury:2024jlz,Firouzjahi:2023ahg,Firouzjahi:2023aum, Iacconi:2023ggt,Davies:2023hhn,Jackson:2023obv,Riotto:2024ayo,Banerjee:2021lqu,Choudhury:2023kam,Choudhury:2024dei,Choudhury:2024dzw,Choudhury:2024aji,Choudhury:2024kjj,Choudhury:2024ezx,Choudhury:2025kxg}. A significant and ongoing dispute has emerged regarding the possibility of producing PBHs from single field inflation in this vast array of methods \cite{Kristiano:2022maq,Kristiano:2023scm,Riotto:2023hoz,Riotto:2023gpm,Firouzjahi:2023aum,Firouzjahi:2023ahg,Firouzjahi:2023bkt,Choudhury:2023vuj,Choudhury:2023jlt,Choudhury:2023rks}.  The main topic of this discussion is whether it is possible to create PBHs in the presence of significant quantum corrections from the small scales during inflation, particularly in the solar-mass domain. The main notion is that in order to generate PBHs, there must be a moment during inflation where the primordial fluctuations see a rapid rise in their power. Having an ultra-slow roll (USR) regime in addition to the standard slow-roll (SR) features during inflation is one of the simplest ways to achieve the aforementioned circumstances. This regime should ideally be brief in order to limit the enhancements. But up until now, the argument has been driven by the need for an appropriate theoretical analysis to examine the effects of quantum loop corrections on the scalar power spectrum and manage the SR to USR transition.

After the matching mode re-enters the Hubble horizon during radiation dominance, the theory of creating black holes from the gravitational collapse of massive primordial disturbances in the early Universe is already rather old. It is even more crucial to carefully examine the underlying physics and devise practical solutions to address this problem in light of the recent debate casting doubt on their formation prospects. One such step toward this approach was taken by the authors of \cite{Choudhury:2023vuj,Choudhury:2023jlt,Choudhury:2023rks}, who introduced a new procedure to correct the power spectrum calculations from any concerning divergences brought up by quantum loop corrections and ultimately produce an expression that represents the two-point correlation function that is physically relevant after one-loop corrections. A no-go theorem that firmly limits the production of any PBHs over $M_{\rm PBH}\sim {\cal O}(10^{2}){\rm gm}$ in single-field inflation models was also derived from the result. The no-go theorem also suggests that the total e-foldings of expansion stop with $\Delta N_{\rm Total}\sim {\cal O}(25)$ if one wants to produce solar-mass, ${\cal O}(M_{\odot})$ PBHs. We find it crucial to develop alternatives that can avoid this strong no-go bound on PBH mass because the above regularization-renormalization-resummation (RRR) procedure that leads to the no-go theorem is developed in a completely model-independent manner, and the renormalization procedure used for the quantum loop corrections is also scheme-independent. Refs.\cite{Choudhury:2023hvf,Choudhury:2023kdb,Choudhury:2023hfm,Choudhury:2023fwk,Choudhury:2024one,Bhattacharya:2023ysp,Choudhury:2023fjs} already have some of the intriguing possibilities.

In this essay, we examine the theory of contraction and non-singular bounce \cite{Khoury:2001wf,Khoury:2001zk,Khoury:2001bz,Buchbinder:2007ad,Lehners:2007ac,Lehners:2008vx,Brandenberger:2012zb,Chowdhury:2015cma,Cai:2011tc,Brandenberger:2016vhg,Boyle:2004gv,Wands:1998yp,Peter:2002cn,Allen:2004vz,Martin:2003sf,Brustein:1998kq,Starobinsky:1980te,Mukhanov:1991zn,Brandenberger:1993ef,Novello:2008ra,Lilley:2015ksa,Battefeld:2014uga,Peter:2008qz,Biswas:2005qr,Bamba:2013fha,Nojiri:2014zqa,Bajardi:2020fxh,Bhargava:2020fhl,Cai:2009in,Cai:2012ag,Shtanov:2002mb,Ilyas:2020qja,Ilyas:2020zcb,Zhu:2021whu,Banerjee:2016hom,Saridakis:2018fth,Barca:2021qdn,Wilson-Ewing:2012lmx,K:2023gsi,Agullo:2020cvg,Agullo:2020fbw,Agullo:2020wur,Stargen:2016cft,Sriramkumar:2015yza,Banerjee:2022gpy,Paul:2022mup,Odintsov:2021yva,Banerjee:2020uil,Das:2017jrl,Pan:2024ydt,Colas:2024xjy,Piao:2003zm,Cai:2017dyi,Cai:2017pga,Cai:2015nya,Cai:2019hge,Zhu:2023lbf} as a new paradigm for studying the production of huge mass PBHs. This is followed by normal inflation in the presence of a USR phase and concluding with another SR phase. In the present context of having contraction and bounce prior to inflation, we seek to give a thorough explanation of the three required processes of regularization, renormalization, and resummation of the quantum loop contributions. Our current approach, which is likewise model-independent, expands the previously established theoretical methods to a considerably wider field of inquiry that goes beyond inflation. The constraint on e-folds that the no-go theorem implies for producing large mass PBHs is the other requirement to satisfy. Therefore, we further aim to demonstrate with the current theory the possibility of achieving the minimum requirement of $N_{\rm Total}\sim {\cal O}(60)$. We observe that the peak spectrum amplitude can lie within $10^{-3}\leq A \leq 10^{-2}$ when the effective sound speed is varied between $0.88\leq c_{s}\leq 1$. This suggests that causality and unitarity are still protected in the theory. Next, using the Press-Schechter or threshold statistics approach, we analyze the production of PBHs and determine their current fraction of energy density contained in dark matter, or abundance, $f_{\rm PBH}$. We highlight the projected mass windows with considerable abundance, $10^{-3}\leq f_{\rm PBH}< 1$, and confront our PBH abundance estimates with the restrictions from microlensing experiments to validate our findings in light of the most recent numerical analysis of the observational data. In order to regulate PBH overproduction, microlensing constraints are applied to PBH abundances across $10^{-33}\!-\!10^{-27}$ and $10^{-6}\!-\!10^{-1}M_\odot$.

\section{The five-phase scenario for EFT of non-singular bounce}\label{sec2}
\begin{figure*}[ht!]
    	\centering
    \subfigure[]{
      	\includegraphics[width=7.9cm,height=5.8cm]{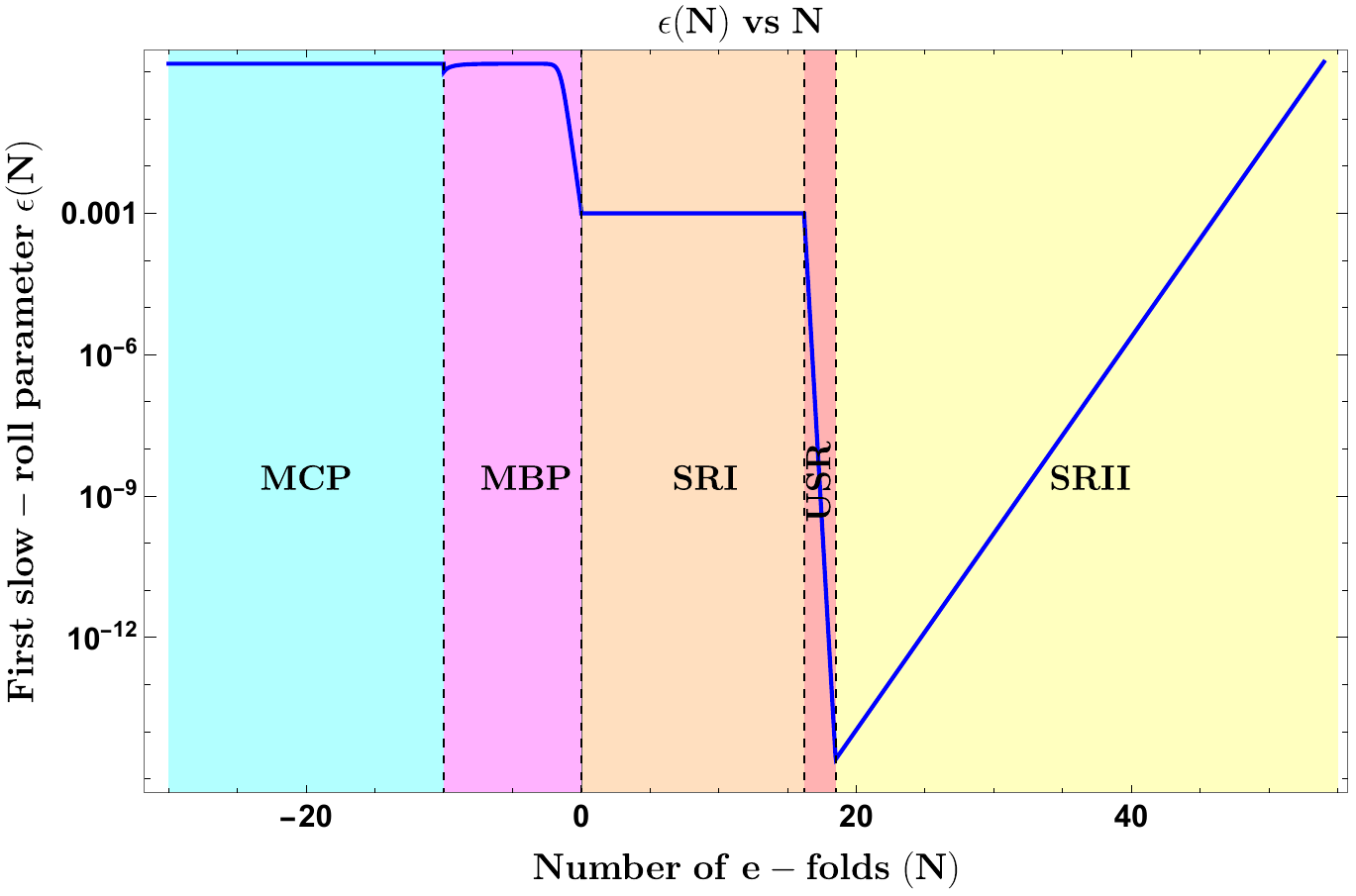}
        \label{epsilon}
    }
    \subfigure[]{
        \includegraphics[width=7.9cm,height=5.8cm]{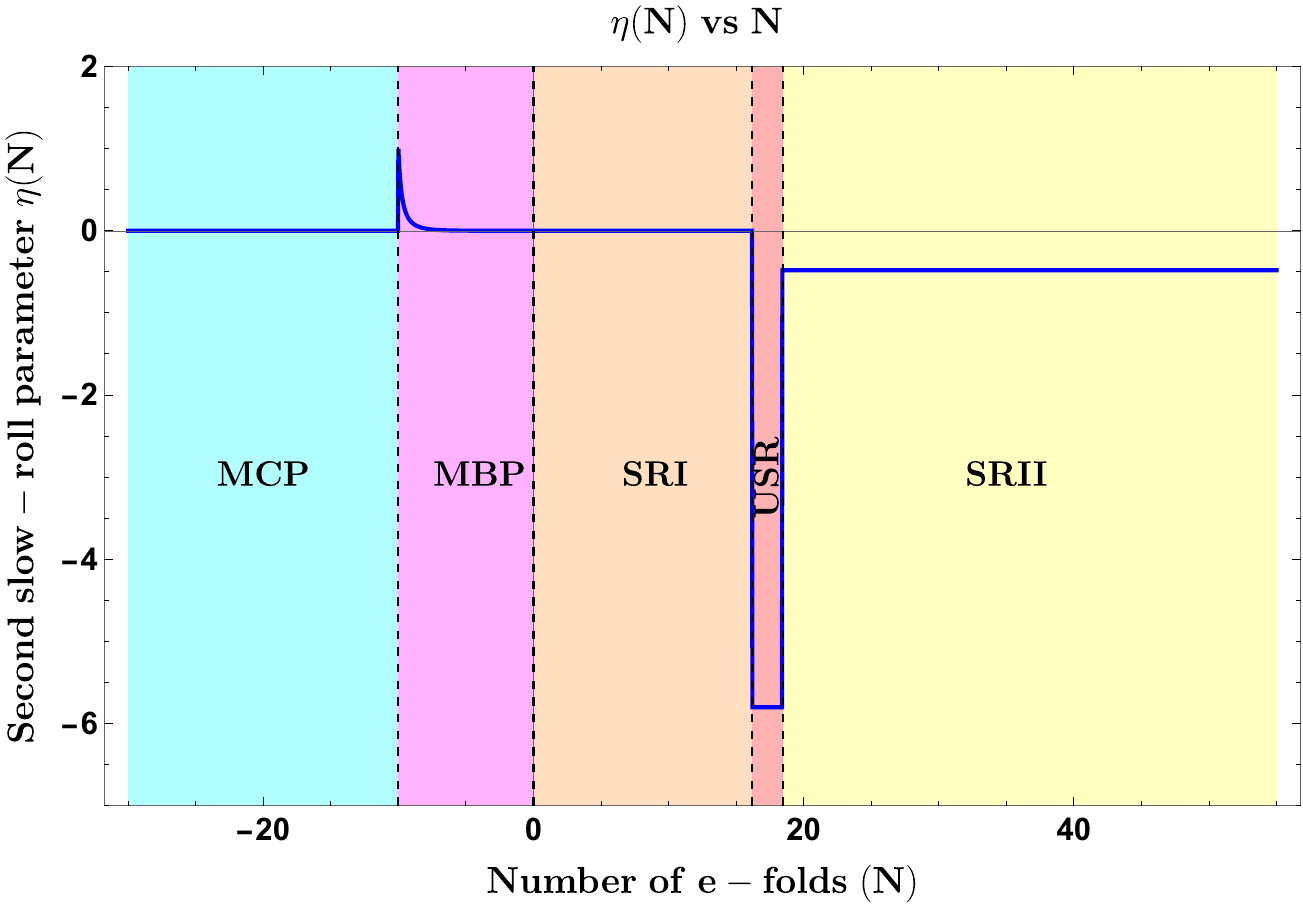}
        \label{eta}
    }
    	\caption[Optional caption for list of figures]{Graphical feature of the \ref{epsilon} $\epsilon({ N})$ and \ref{eta} $\eta({ N})$ in the presence of matter contraction and bouncing phase along with a USR phase as a function of $N$ for sharp transitions.} 
    	\label{dyn}
    \end{figure*}

\textcolor{black}{The main objective of the EFT of Inflation is to provide a model-independent, symmetry-based description of inflationary dynamics when the UV completion is unknown.  
It exploits the spontaneous breaking of time-translation symmetry by the inflationary background to write the most general action allowed by the remaining symmetries.  
EFT isolates the relevant low-energy degrees of freedom (the propagating fluctuations) without committing to a specific scalar potential.  
It systematically organizes all operators by relevance, using a controlled expansion in energy scales and derivatives.  
Unknown high-energy effects enter as higher-dimension operators suppressed by a cutoff, making their impact parametrizable.  
These operators can change sound speed, generate non-Gaussianity, and modify the tensor-to-scalar ratio and power spectrum.  
The framework unifies many seemingly different inflationary models under common effective parameters, easing comparison with data.  
EFT provides clear stability criteria (no ghosts, gradients, or tachyons) via conditions on operator coefficients.  
It links theoretical predictions directly to observables measured in the CMB and large-scale structure.  
EFT organizes perturbative calculations reliably and permits consistent treatment of non-renormalizable interactions.  
Overall, it bridges unknown microscopic physics and measurable cosmological signatures in a systematic, testable way.} The framework begins with a model-independent effective field theory (EFT) action valid below a UV cutoff, where symmetry principles constrain the allowed operators and enable bounds on parameters such as the speed of sound ($c_s$) in terms of EFT coefficients \cite{Weinberg:2008hq,Cheung:2007st,Choudhury:2017glj,Delacretaz:2016nhw}. Let us start with the following action:
\bea
 S_{\rm EFT}&=&\displaystyle\int d^{4}x \sqrt{-g}\left[\frac{M^2_{pl}}{2}R+M^2_{pl} \dot{H} g^{00}-M^2_{pl} \left(3H^2+\dot{H}\right)+{\cal W}\left(\delta g^{00}, \delta K^{\mu\nu},\cdots\right)\right].
	\eea
In this case, the final term ${\cal W}\left(\delta g^{00}, \delta K^{\mu\nu},\cdots\right)$ physically represents all potential contributions from the tiny FLRW background fluctuations, described by the scale factors, I. $a(\tau)=a_0(\tau/\tau_0)^\alpha$ and II. $a(\tau)=a_0[1+(\tau/\tau_0)^2]^{2\alpha}$ with $\alpha=1/(\epsilon-1)$~\footnote{\textcolor{black}{Henceforth in this paper we will use the conformal time coordinate and the number of e-foldings, which can be expressed in terms of the cosmic physical time by, $\tau=\int dt/a(t)$ and $N=\int Hdt$. relationship between the conformal time and number of e-folds hence can be written as, $\tau=\int dN/{\cal H}$ where ${\cal H}=a^{'}/a=Ha=\dot{a}$.}}. For the first scale factor $\epsilon=3/2$ and $\epsilon>3$ correspond to matter and ekpyrotic contraction. On the other hand, for the second scale factor $\epsilon=3/2$ and $\epsilon>3$ correspond to matter and ekpyrotic bounce. Here, $K_{\mu\nu}$ defines the extrinsic curvature and $H$ is the Hubble parameter.  The Goldstone mode ($\pi$), which is connected to the curvature perturbation variable ($\zeta$) via the linear mapping in the associated second-order perturbed EFT action, can be used to further characterize the imprints of the scalar perturbations at the comoving scale. Following the establishment of this connection, a second-order differential equation known as the Mukhanov Sasaki equation is used to describe the curvature perturbation variable ($\zeta$), which physically describes a system of oscillators with a time-dependent frequency in momentum space.
  
   This article focuses on the five-phase scenario, which captures all the necessary ingredients to generate large amplitude fluctuations in the presence of EFT of non-singular bounce. In more specific terms, this can be accomplished by considering an ekpyrotic/matter contraction and bounce, followed by an ultra-slow roll (USR) phase sandwiched between two slow roll (SRI and SRII) stages. At the end of the SRII phase, inflation ceases when the number of e-foldings $N$ hits the magic number $60$, which is required to achieve inflation. Additionally, the second slow-roll parameter can be further parameterized in the following simple manner based on the behavior of the transition from different phases: 
   \bea
\eta(N) &=&  \eta_{\rm C}(N\leq N_{c})+ \eta_{\rm B}(N\leq N_{b})+\eta_{\rm SRI}(N_b<N\leq N_{s}) \nonumber\\
&&~~~~~~~+\Theta\left(N-N_s\right)\eta_{\rm USR}(N_{s}\leq N\leq N_{e}) + \Theta\left(N-N_e\right)\eta_{\rm SRII}(N_{e}\leq N\leq N_{\rm end}).\quad
\eea
It is to be noted that, $\eta_{\rm c},\;\eta_{\rm b},\;\eta_{\rm SRI},\;\eta_{\rm USR},\;\eta_{\rm SRII}$ characterizes constants appearing in contraction, bouncing, SRI, USR, and SRII periods respectively. Utilizing this mentioned structure of $\eta$ one can further construct the form of the first slow-roll parameter, $\epsilon(N) =-d\ln H/dN$. In figure \ref{epsilon} and \ref{eta} we have depicted the features of the first and second slow-roll parameters with respect to the number of e-foldings $N$ in the presence of matter contraction and bouncing phase along with the sharp transitions in the SRI to USR and USR to SRII boundaries. Almost similar features in the graphical plots can be visualized in the case of ekpyrotic contraction and bounce.

\textcolor{black}{Further this discussion focuses on determining the solutions for the comoving curvature perturbation in a spatially flat FLRW background using the underlying Goldstone EFT framework. Ekpyrotic/matter contraction, ekpyrotic/matter bounce, first slow roll (SRI), ultra slow roll (USR), and, finally, the second slow roll phase (SRII) make up our configuration. To safely examine the behavior of the mode solutions in the five phases, we employ the decoupling limit. Finding relationships between the comoving curvature perturbation and its conjugate momentum variable will then help create the various components of the scalar modes' power spectrum. The solutions to the Mukhanov-Sasaki equation for each of the five previously discussed phases are examined in this section. The answers discovered will be essential for our analysis of the power spectrum at the tree and loop levels in the current context of our inquiry. The curvature perturbation modes for different phases may be obtained by solving the Mukhanov-Sasaki (MS) equation in Fourier space after getting it by modification of the EFT action. The following is the expression for the Fourier space MS equation:}
\bea \label{MSfourier}
\bigg[\frac{d^2}{d\tau^2}+2 \frac{z'(\tau)}{z(\tau)}\partial_{\tau}+c_s ^2 k^2\bigg]\zeta_{\bf k}(\tau) = 0.
\eea 
\textcolor{black}{The following statement contains the {\it Mukhanov-Sasaki} variable, which in this instance is represented as $z(\tau)$:}
\bea z(\tau):=\frac{a(\tau)\sqrt{2\epsilon}}{c_s}.\eea
\textcolor{black}{The presence of the aforementioned five successive phases may be readily explained by utilizing the explicit definition of the conformal time-dependent scale factor $a(\tau)$. The initial slow roll parameter $\epsilon$ takes distinct values throughout different phases, and its conformal time-dependent behavior also turns out to be different in each of these five phases. This is important to notice in order to prevent additional confusion. This is really important information that will be used a lot in the rest of this work's analysis. In order to solve the second-order differential equation mentioned above, we now use the following finding:}
\bea  \frac{z'(\tau)}{z(\tau)}={\cal H}\left(1-\eta+\epsilon-s\right)\quad\quad\quad{\rm where}\quad\quad\quad s:=\frac{c^{'}_s}{{\cal H}c_s}.\eea
\textcolor{black}{We will also introduce another parameter, $\nu$, which is defined by the following formula and occasionally designated as an effective mass parameter in the relevant contexts:}
\bea \nu:\equiv \frac{1}{2}-\frac{1}{\epsilon-1}+\frac{\eta}{\epsilon-1}-\frac{3s}{\epsilon-1}.\eea
\textcolor{black}{Now let's talk about the ramifications of these newly specified mass parameters in the next five stages, which are point-by-point attached below:}
\begin{itemize}
    \item \textcolor{black}{The first SR parameter is extremely tiny in the inflationary phase, i.e., for SRI, USR, and SRII phases, where $\epsilon\ll 1$. The following simplified formulation describes the effective mass parameter in certain situations:}
     \bea \nu\approx\frac{3}{2}+\epsilon-\eta-3s.\eea

     \item \textcolor{black}{The initial slow-roll parameter in the matter contraction and bouncing situations is $\epsilon=3/2$. In these situations, the simplified statement that follows describes the effective mass parameter:}
          \bea \nu= \frac{1}{2}+2+2\eta+6s=\frac{5}{2}+2\eta+6s.\eea

          \item \textcolor{black}{The initial SR parameter in the ekpyrotic contraction and bounce scenarios must be $\epsilon>3$, that is, $\epsilon=7/2(=3.5)$. As a result, the following simplified statement describes the effective mass parameter:}
          \bea \nu= \frac{1}{2}+\frac{2}{5}+\frac{2}{5}\eta+\frac{6}{5}s=\frac{9}{10}+\frac{2}{5}\eta+\frac{6}{5}s.\eea
\end{itemize}
The generic solution for the curvature perturbation modes in momentum space from the {\it Mukhanov Sasaki equation} can be further expressed as follows by using the previously given facts shown in the sequential five-phase scenario:
\bea
 &&\zeta_{\bf k}(\tau)=\left(\frac{2^{\nu-\frac{3}{2}}c_sH }{2iM_{ pl}\sqrt{\epsilon_{P}(\tau)}}\right)\frac{(-k c_s \tau )^{\frac{3}{2}-\nu}}{(c_sk)^{3/2}}\Bigg|\frac{\Gamma(\nu)}{\Gamma(\frac{3}{2})}\Bigg |\bigg[\alpha^{(Q)}_{{\bf k}}\left(1+ikc_s\tau\right)\; e^{-i\left(k c_s\tau+\frac{\pi}{2}\left(\nu+\frac{1}{2}\right)\right)}\nonumber\\
 &&~~~~~~~~~~~~~~~~~~~~~~~~~~~~~~~~~~~~~~~~~~~~~~~~~~~~~-\beta^{(Q)}_{{\bf k}}\left(1-ikc_s\tau\right)\; e^{i\left(k c_s\tau+\frac{\pi}{2}\left(\nu+\frac{1}{2}\right)\right)}\bigg]\forall Q,\quad\quad\quad
 \eea
where $Q$ is given by $Q= 1 ({\rm Contraction})$, $2 ({\rm Bounce})$, $3 ({\rm SRI})$, $4 ({\rm USR})$, $5 ({\rm SRII})$ for the five phases. Here we use the following facts:
\bea &&\epsilon_{\rm C}=\epsilon_{\rm SRI}\left(a^2_0\tau^2\left(\frac{\tau}{\tau_0}\right)^{2\alpha}\frac{\epsilon_{\rm c,0}}{\epsilon_{\rm SRI}}\right)\textcolor{black}{=\epsilon_{\rm SRI}\left(a^2_0\frac{N^2}{{\cal H}^2}\left(\frac{{\cal H}_0}{\cal H}\right)^{2\alpha}\left(\frac{N}{N_0}\right)^{2\alpha}\frac{\epsilon_{\rm c,0}}{\epsilon_{\rm SRI}}\right)}, \\
&&\epsilon_{\rm B}=\epsilon_{\rm SRI}\left(a^2_0\tau^2\left[1+\left(\frac{\tau}{\tau_0}\right)^{2}\right]^{\alpha}\frac{\epsilon_{\rm b,0}}{\epsilon_{\rm SRI}}\right)\textcolor{black}{=\epsilon_{\rm SRI}\left(a^2_0\frac{N^2}{{\cal H}^2}\left[1+\left(\frac{{\cal H}_0}{\cal H}\right)^{2}\left(\frac{N}{N_0}\right)^{2}\right]^{\alpha}\frac{\epsilon_{\rm b,0}}{\epsilon_{\rm SRI}}\right)},~~~~\\
&&\textcolor{black}{\epsilon_{\rm SRI}=1-\frac{{\cal H}^{'}}{{\cal H}^2}=1-\frac{1}{\cal H}\left(\frac{d{\cal H}}{dN}\right)},\\
&&\epsilon_{\rm USR}=\epsilon_{\rm SRI}\left(\frac{\tau_s}{\tau}\right)^6\textcolor{black}{=\epsilon_{\rm SRI}\left(\frac{{\cal H}}{{\cal H}_s}\right)^6\left(\frac{N_s}{N}\right)^6},\\
&&\epsilon_{\rm SRII}=\epsilon_{\rm SRI}\left(\frac{\tau_s}{\tau_e}\right)^6\textcolor{black}{=\epsilon_{\rm SRI}\left(\frac{{\cal H}_e}{{\cal H}_s}\right)^6\left(\frac{N_s}{N_e}\right)^6}.\eea 
Here it is important to note that, $\alpha^{(Q)}_{{\bf k}}$ and $\beta^{(Q)}_{{\bf k}}$ are the Bogoliubov coefficients for five consecutive periods. By applying the continuity and differentiability conditions of the scalar modes, one can further fix the expressions for both Bogoliubov coefficients in the USR and SRII phases for a given initial choice of vacuum, which is typically described in terms of the {\it Bunch Davies} states in the contraction, bouncing, and SRI phase. See ref. \cite{Choudhury:2024dei} for more details on the structure of the Bogoliubov coefficients. 

\section{Tree level spectrum for large fluctuations}
\begin{figure*}[htb!]
    	\centering
    \subfigure[]{
      	\includegraphics[width=7.9cm,height=5.8cm]{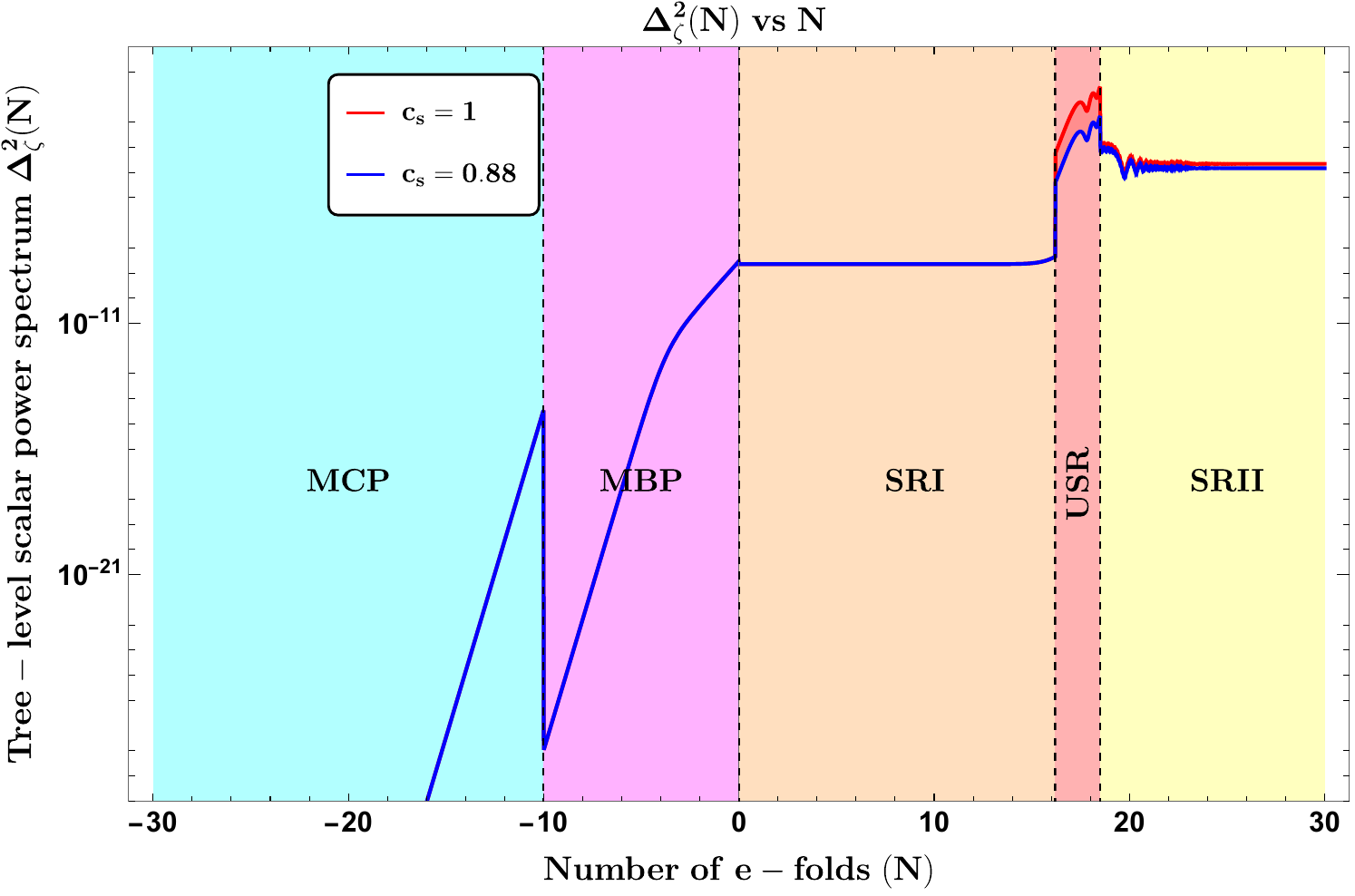}
        \label{treematter}
    }
    \subfigure[]{
       \includegraphics[width=7.9cm,height=5.8cm]{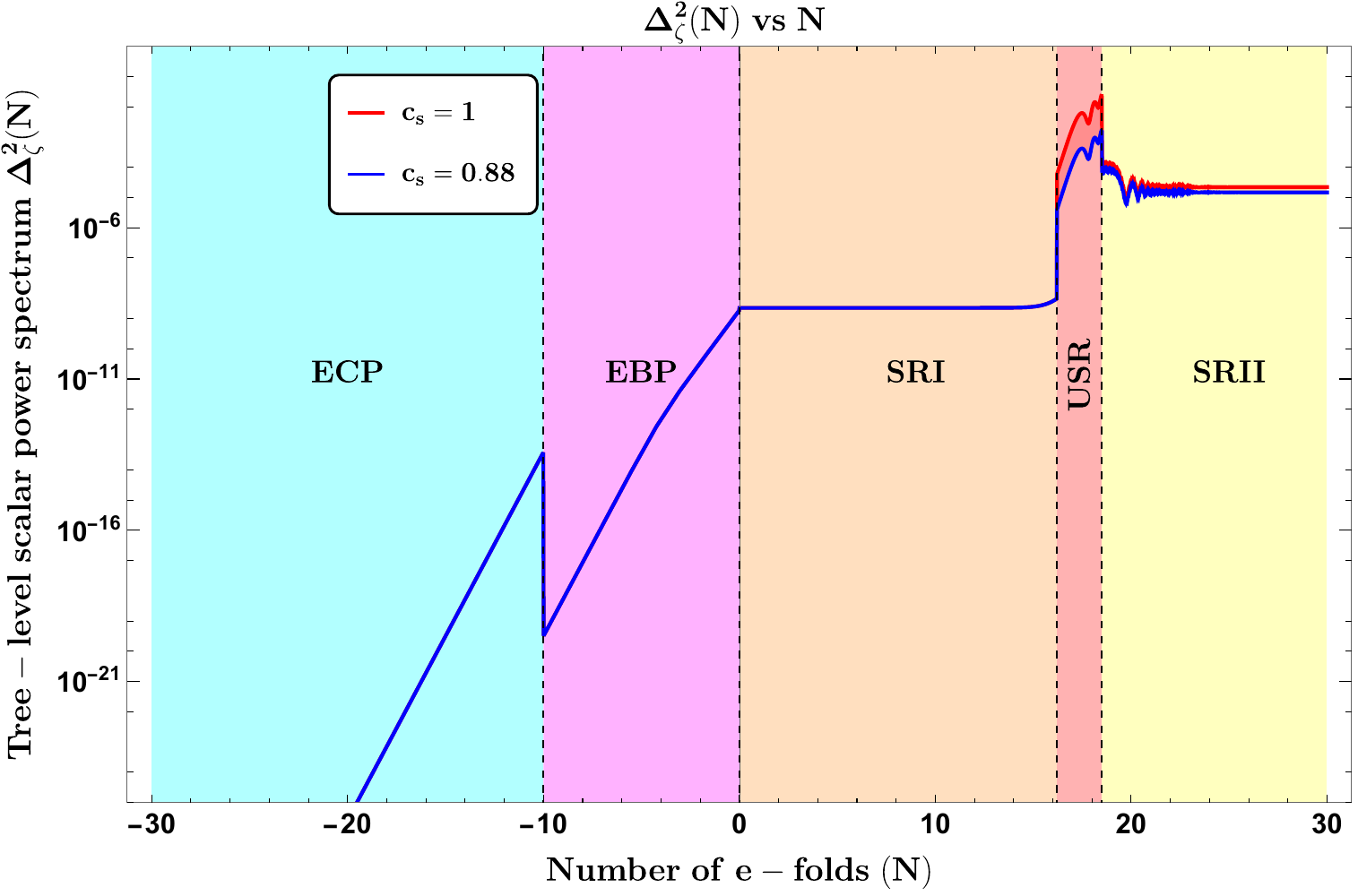}
        \label{treeekpy}
    }
    	\caption[Optional caption for list of figures]{Plots of tree-level scalar power spectrum as a function of the e-foldings ${\rm N}$. In the left we observe the matter contraction (MCP) and bounce (MBP) phases where $\epsilon=3/2$. In the right we observe the ekpyrotic contraction (ECP) and bounce (EBP) phases where $\epsilon=7/2$.  } 
    	\label{treelevelpspec }
    \end{figure*}
The tree-level power spectrum, which describes the large fluctuations and PBH formation for the previously mentioned five-phase scenraio (in the presence of contraction and bouncing phase) is represented by the following expression: 
\bea \label{tree}\Delta^{2}_{\zeta,{\bf Tree}}(k)&=&\Delta^{2}_{\zeta,{\bf SRI}}(k)\times\Bigg\{1+\left(\frac{\epsilon_{\rm SRI}}{\epsilon_c}\right)\left(\frac{k}{k_*}\right)^{\frac{2\epsilon_c}{\epsilon_c-1}}+\left(\frac{\epsilon_{\rm SRI}}{\epsilon_b}\right)\left(\frac{k}{k_*}\right)^{2}\left[1+\left(\frac{k_*}{k}\right)^2\right]^{-\frac{1}{(\epsilon_b-1)}}\nonumber\\
&&\quad\quad\quad\quad+\left(\frac{k_e}{k_s }\right)^{6}\bigg[\Theta(k-k_s)\left|\alpha^{(4)}_{\bf k}-\beta^{(4)}_{\bf k}\right|^{2}+\Theta(k-k_e)\left|\alpha^{(5)}_{\bf k}-\beta^{(5)}_{\bf k}\right|^{2}\bigg]\Bigg\}.\quad\quad\eea
The power spectrum in the SRI phase is represented by the first term in the expression above, which is provided by, \bea \Delta^{2}_{\zeta,{\bf SRI}}(k)=\left(\frac{2^{2\nu-3}H^{2}}{8\pi^{2}M^{2}_{ pl}\epsilon c_s}\left|\frac{\Gamma(\nu)}{\Gamma\left(\frac{3}{2}\right)}\right|^2\right)_*.\eea It is significant to notice that the pivot scale is used to quantify the expression for the power spectrum in the SRI phase. The tree-level spectrum with respect to the e-foldings is depicted in figure \ref{treematter} and \ref{treeekpy}.

\section{RRR loop-level spectrum for large fluctuations}

In order to further explain the quantum loop effects, let us examine the bare interaction term \bea H^{\bf B}_{\rm int}(\tau)=\int d^3x~ M^2_{pl}a^2~\frac{1}{2c^2_s}\epsilon\partial_{\tau}\left(\frac{\eta}{c^2_s}\right)\zeta^{'}\zeta^{2},\eea that appears in the third-order perturbed action for the scalar modes. This term directly affects the computation at the one-loop level. The influence of this contribution can be calculated directly using the well-known Schwinger-Keldysh formalism, which is given by: 
\bea \label{Hamx}\langle\hat{\zeta}_{\bf p}\hat{\zeta}_{-{\bf p}}\rangle:&=&\lim_{\tau\rightarrow 0}\left\langle\bigg[\overline{T}e^{i\int^{\tau}_{-\infty(1-i\epsilon)}d\tau^{'}\;H_{\rm int}(\tau^{'})}\bigg]\;\;\hat{\zeta}_{\bf p}(\tau)\hat{\zeta}_{-{\bf p}}(\tau)
\;\;\bigg[{T}e^{-i\int^{\tau}_{-\infty(1+i\epsilon)}d\tau^{''}\;H_{\rm int}(\tau^{''})}\bigg]\right\rangle\nonumber\\
&=&\underbrace{\langle\hat{\zeta}_{\bf p}\hat{\zeta}_{-{\bf p}}\rangle_{(0,0)}}_{\bf Tree}+\underbrace{\langle\hat{\zeta}_{\bf p}\hat{\zeta}_{-{\bf p}}\rangle_{(0,1)}+\langle\hat{\zeta}_{\bf p}\hat{\zeta}_{-{\bf p}}\rangle^{\dagger}_{(0,1)}+\langle\hat{\zeta}_{\bf p}\hat{\zeta}_{-{\bf p}}\rangle_{(0,2)}+\langle\hat{\zeta}_{\bf p}\hat{\zeta}_{-{\bf p}}\rangle^{\dagger}_{(0,2)}+\langle\hat{\zeta}_{\bf p}\hat{\zeta}_{-{\bf p}}\rangle_{(1,1)}}_{\bf One-loop},\quad\quad \eea
\textcolor{black}{The following discussion contains explicit contributions that occur in the tree level and the one-loop level result of the two point correlation function of the scalar modes, which we must assess for the five successive phases in this work that were previously mentioned:}
\bea
     &&\label{c0}\langle\hat{\zeta}_{\bf p}\hat{\zeta}_{-{\bf p}}\rangle_{(0,0)}=\lim_{\tau\rightarrow 0}\langle \hat{\zeta}_{\bf p}(\tau)\hat{\zeta}_{-{\bf p}}(\tau)\rangle,\\
    &&\label{c1}\langle\hat{\zeta}_{\bf p}\hat{\zeta}_{-{\bf p}}\rangle_{(0,1)}=-i\lim_{\tau\rightarrow 0}\int^{\tau}_{-\infty}d\tau_1\;\langle \hat{\zeta}_{\bf p}(\tau)\hat{\zeta}_{-{\bf p}}(\tau)H_{\rm int}(\tau_1)\rangle,\\
 &&\label{c2}\langle\hat{\zeta}_{\bf p}\hat{\zeta}_{-{\bf p}}\rangle^{\dagger}_{(0,1)}=-i\lim_{\tau\rightarrow 0}\int^{\tau}_{-\infty}d\tau_1\;\langle \hat{\zeta}_{\bf p}(\tau)\hat{\zeta}_{-{\bf p}}(\tau)H_{\rm int}(\tau_1)\rangle^{\dagger},\\
 &&\label{c3}\langle\hat{\zeta}_{\bf p}\hat{\zeta}_{-{\bf p}}\rangle_{(0,2)}=\lim_{\tau\rightarrow 0}\int^{\tau}_{-\infty}d\tau_1\;\int^{\tau}_{-\infty}d\tau_2\;\langle \hat{\zeta}_{\bf p}(\tau)\hat{\zeta}_{-{\bf p}}(\tau)H_{\rm int}(\tau_1)H_{\rm int}(\tau_2)\rangle,\\
 &&\label{c4}\langle\hat{\zeta}_{\bf p}\hat{\zeta}_{-{\bf p}}\rangle^{\dagger}_{(0,2)}=\lim_{\tau\rightarrow 0}\int^{\tau}_{-\infty}d\tau_1\;\int^{\tau}_{-\infty}d\tau_2\;\langle \hat{\zeta}_{\bf p}(\tau)\hat{\zeta}_{-{\bf p}}(\tau)H_{\rm int}(\tau_1)H_{\rm int}(\tau_2)\rangle^{\dagger},\\
  &&\label{c5}\langle\hat{\zeta}_{\bf p}\hat{\zeta}_{-{\bf p}}\rangle^{\dagger}_{(1,1)}=\lim_{\tau\rightarrow 0}\int^{\tau}_{-\infty}d\tau_1\;\int^{\tau}_{-\infty}d\tau_2\;\langle H_{\rm int}(\tau_1)\hat{\zeta}_{\bf p}(\tau)\hat{\zeta}_{-{\bf p}}(\tau)H_{\rm int}(\tau_2)\rangle^{\dagger}.\eea

\textcolor{black}{For the detailed derivation we refer to the references \cite{Choudhury:2024dei,Choudhury:2023hvf,Choudhury:2023rks,Choudhury:2023jlt,Choudhury:2023vuj}, where all the relevant details can be found.} Consequently, we get the following contribution after implementing cut-off regularization in the momentum integrations appearing in each of the previously mentioned five-phases:
\bea \label{one-loopR} \Delta^{2}_{\zeta, {\bf R}}(k)&=&\bigg[\Delta^{2}_{\zeta,{\bf Tree}}(k)\bigg]_{\bf SRI}\nonumber\\
&&~~~~~~\times\bigg(1+\bigg[\Delta^{2}_{\zeta,{\bf Tree}}(k)\bigg]_{\bf SRI}\Bigg\{\underbrace{{\bf W}_{\bf C}+{\bf W}_{\bf B}+{\bf W}_{\bf SRI}+{\bf W}_{\bf USR}+{\bf W}_{\bf SRII}}_{\textbf{Regularized one-loop correction}}\Bigg\}\bigg).~~~~~~~~\eea
Here ${\bf W}_{\bf C}$, ${\bf W}_{\bf B}$, ${\bf W}_{\bf SRI}$, ${\bf W}_{\bf USR}$ and ${\bf W}_{\bf SRII}$ are the one-loop contribution from the contraction, bouncing, SRI, USR and SRII phases which have quadratic UV divergent term $\left(\Lambda/H\right)^2$ and the logarithmic divergent term $\ln\left(\Lambda/H\right)$~\footnote{Here $\Lambda$ represents the cut-off of the regularization appearing in the momentum loop integrals. UV divergences are the more detrimental of the two forms of divergences in this situation, because they directly call into question the application of perturbation theory.}, are described in the ref \cite{Choudhury:2024dei}.

We must use the renormalization procedure in this situation to eliminate the UV and IR divergent contributions after correctly completing the regularization and retrieving them. In theory, we can include the counter-terms with an appropriate renormalization technique. Here we need to use:
\bea \zeta_{\bf R}=\zeta-\zeta_{\bf C}=\sqrt{{\bf Z}^{\rm IR}}\times \zeta_{\bf B}\quad\quad\quad{\rm where}\quad\quad\quad {\bf Z}^{\rm IR}:=\left(1+\delta_{{\bf Z}^{\rm IR}}\right).\eea
Here $\zeta_{\bf R}$, $\zeta$ and $\zeta_{\bf C}$ represent renormalized, bare, and counter-term contributions, respectively. Notably, the number ${\bf Z}^{\rm IR}$, or more specifically $\delta_{{\bf Z}^{\rm IR}}$, is often called the counter-term. This quantity must be explicitly estimated in this computation using the physical renormalization condition. Consequntly, the renormalized version of the previously mentioned interaction can be written as:
\bea H^{\bf R}_{\rm int}(\tau)\approx\int d^3x~ M^2_{pl}a^2~\left(1+\frac{3}{2}\delta_{{\bf Z}^{\rm IR}}+\cdots\right) ~\frac{1}{2c^2_s}\epsilon\partial_{\tau}\left(\frac{\eta}{c^2_s}\right)\zeta^{'}\zeta^{2},\eea 
This immediately leads to the renormalized one-loop corrected primordial scalar power spectrum being totally free of the quadratic UV divergence. Additionally, after the renormalization condition is applied, the less detrimental and softened logarithmic IR divergent contribution emerges with higher powers. Here we have the following contribution:
 \bea \label{one-loopRR} {\Delta^{2}_{\zeta, {\bf RR}}(k)}  
   &=&\bigg[\Delta^{2}_{\zeta,{\bf Tree}}(k)\bigg]_{\bf SRI}\nonumber\\
   &&\times\bigg(1+\bigg[\Delta^{2}_{\zeta,{\bf Tree}}(k)\bigg]_{\bf SRI}\bigg\{\underbrace{\overline{{\bf W}}_{\bf C}+\overline{{\bf W}}_{\bf B}+\overline{{\bf W}}_{\bf SRI}+\overline{{\bf W}}_{\bf USR}+\overline{{\bf W}}_{\bf SRII}}_{\textbf{Regularized+Renormalized one-loop correction}}\bigg\}\bigg).~~~\eea
Here $\overline{\bf W}_{\bf C}$, $\overline{\bf W}_{\bf B}$, $\overline{\bf W}_{\bf SRI}$, $\overline{\bf W}_{\bf USR}$ and $\overline{\bf W}_{\bf SRII}$ are the one-loop renormalized contribution from the contraction, bouncing, SRI, USR and SRII phases which are free from the quadratic UV divergent term and contain the softening version of the logarithmic term along with its higher powers. These terms are explicitly mentioned in the ref. \cite{Choudhury:2024dei}.

\begin{figure*}[htb!]
    	\centering
    \subfigure[]{
      	\includegraphics[width=7.9cm,height=5.8cm]{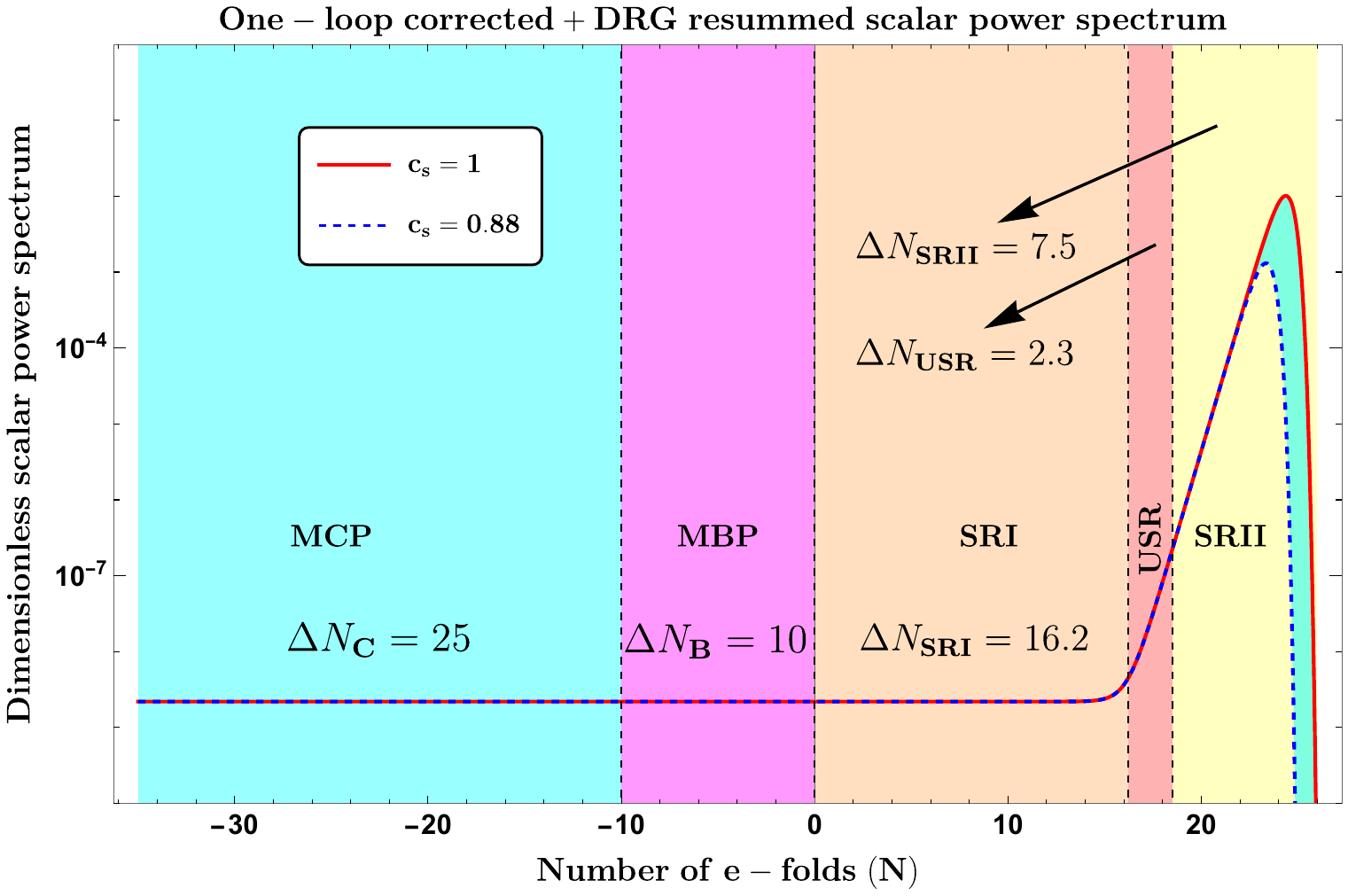}
        \label{drgmatter}
    }
    \subfigure[]{
       \includegraphics[width=7.9cm,height=5.8cm]{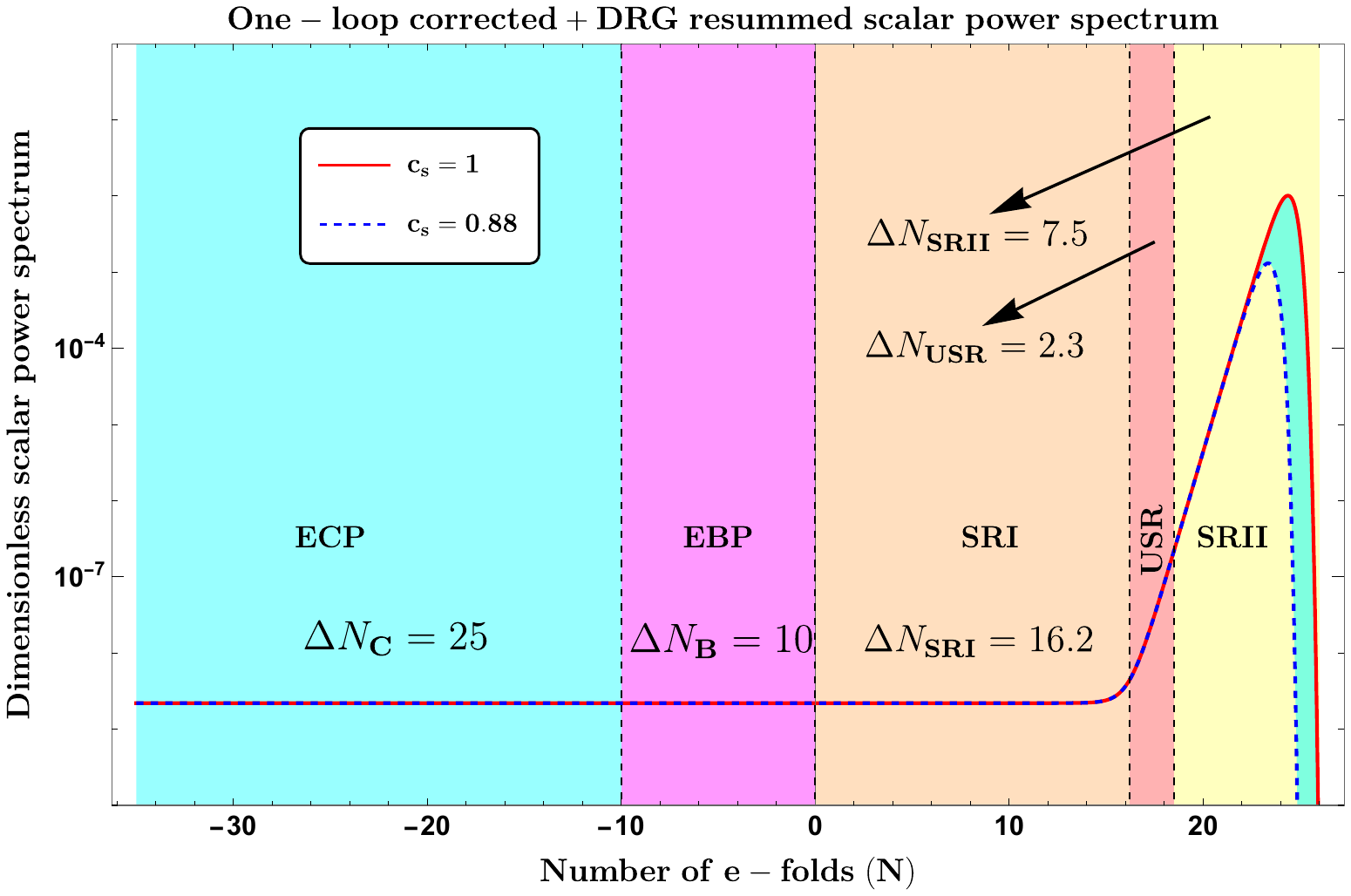}
        \label{drgekpy}
    }
    	\caption[Optional caption for list of figures]{Behaviours of regularized-renormalized-resummed (RRR) scalar power spectrum as a function of the e-foldings ${\rm N}$. The left power spectrum contains the matter contraction (MCP) and bounce (MBP) phases with $\epsilon=3/2$. The right power spectrum contains the ekpyrotic contraction (ECP) and bounce (EBP) phases with $\epsilon=7/2$. } 
    	\label{drgpspec }
    \end{figure*}

Finally, we employ the Dynamical Renormalization Group (DRG) technique \cite{Chen:2016nrs,Baumann:2019ghk,Boyanovsky:1998aa,Boyanovsky:2001ty,Boyanovsky:2003ui,Burgess:2015ajz,Burgess:2014eoa,Burgess:2009bs,Dias:2012qy,Chaykov:2022zro,Chaykov:2022pwd}, also known as exponentiation-influenced resummation, to resum across all potential logarithmic higher-order contributions using the computed renormalized primordial power spectrum for scalar modes. This implies that the DRG approach allows for the sum of all repetitive loop diagrams, and the sum of all conceivable terms in the infinite series yields a finite result due to strict convergence at the scale where the renormalization condition is applied. The most significant effect of the DRG resummed version of the one-loop corrected power spectrum is the resulting coarse-grained form of the power spectrum, which is ultimately quantified as having more softened logarithmic contributions than the renormalized version of the one-loop spectrum:
  \bea &&\Delta^2_{\zeta,{\bf RRR}}=\bigg[\Delta^{2}_{\zeta,{\bf Tree}}(k)\bigg]_{\bf SRI}\nonumber\\
  &&~~~~~~~~~~\times\exp\bigg(
\includegraphics[width=10.9cm,height=1.9cm]{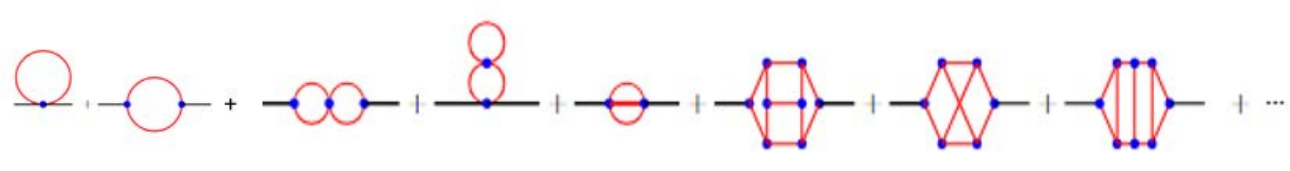}
\bigg).\eea
\textcolor{black}{For the more details on the calculation of DRG resummation scheme please consider the refrences \cite{Choudhury:2024dei,Choudhury:2023hvf,Choudhury:2023rks,Choudhury:2023jlt,Choudhury:2023vuj} for the further details. }
We have calculated the total of all Feynman graphs that appear at every loop level inside the exponential. Figures \ref{drgmatter} and \ref{drgekpy} show the final result of the resummed spectrum with regard to the e-foldings.

\begin{figure*}[ht!]
    	\centering
    {
   \includegraphics[width=10cm,height=6cm]{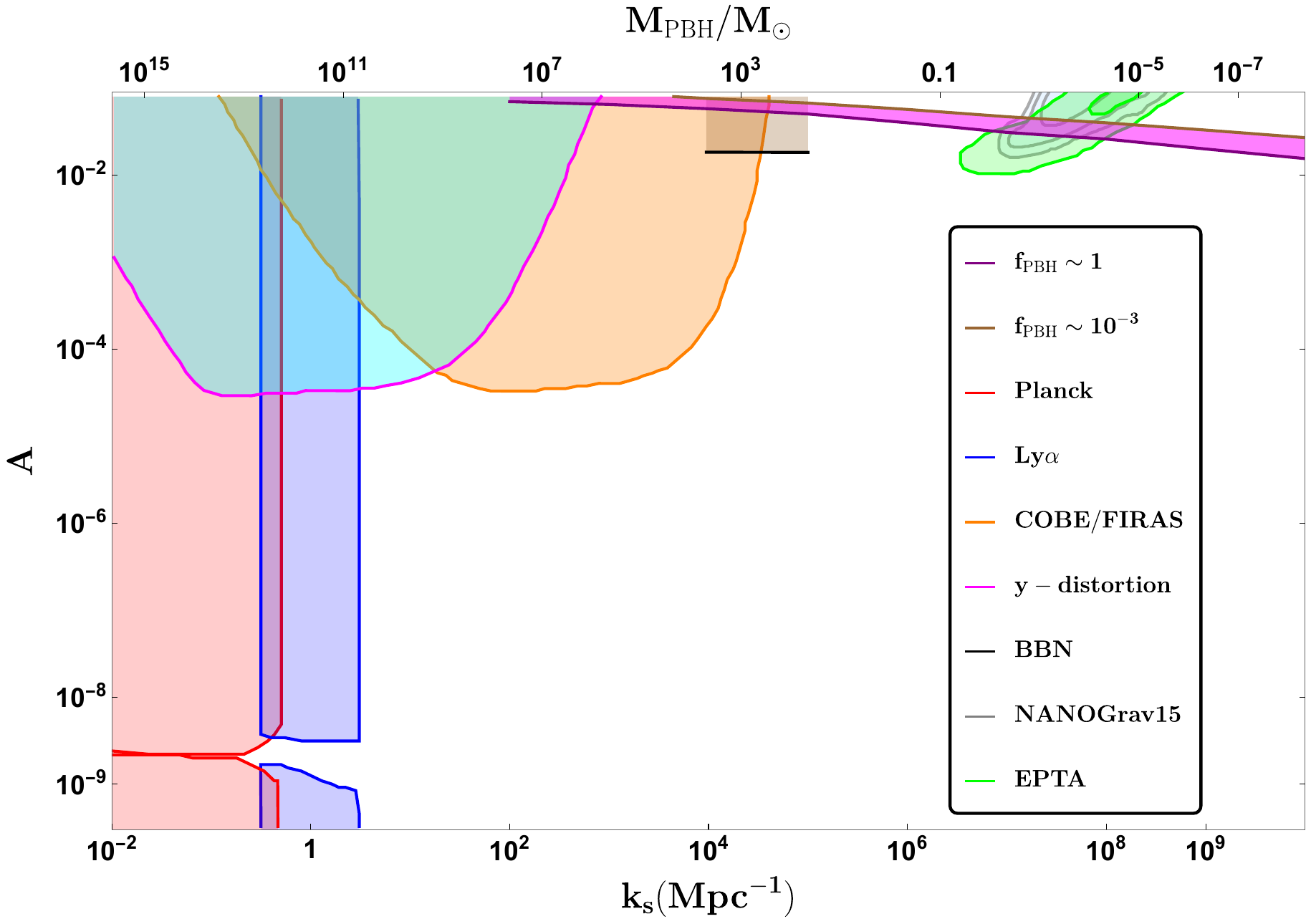}
        \label{distortionbounce}
    } 
    \caption[Optional caption for list of figures]{Scalar power spectrum amplitude necessary to achieve significant PBH abundance as a function of the transition wavenumber $k_{s}$ associated with the PBH mass.  }
\label{amplitudedistortion}
    \end{figure*}

\section{Constraints from PBH formation}
The investigation of substantial boosts in the primordial density perturbations is strongly related to the development of PBHs. One way to provide the required high level of excitement to the perturbations in the early cosmos is to incorporate a USR phase into our arrangement. These huge curvature perturbations experience gravitational instability upon re-entry into the horizon, which causes them to collapse and form PBH after they surpass specific conditions on their formation threshold. For the PBH creation, we intend to perform our current analysis using the Press-Schechter formalism (threshold statistics). The PBH formation threshold is thoroughly examined in this linear regime approximation, yielding the most advantageous range of $2/5\leq \delta_{\rm th}\leq 2/3$. Here the PBHs abundance follow the formula:
\bea \label{fpbhformula}
f_{\rm PBH} = 1.68\times 10^{8}\;\bigg(\frac{\gamma}{0.2}\bigg)^{1/2} \bigg(\frac{g_{*}}{106.75}\bigg)^{-1/4} \bigg(\frac{M_{\rm PBH}}{M_{\odot}}\bigg)^{-1/2} \beta(M_{\rm PBH})~~~{\rm where}~~~\frac{M_{\rm PBH}}{M_{\odot}}\propto \bigg(\frac{k_{*}}{k_{s}}\bigg)^{2},~~~~~~
\eea
where $k_s$ is the transition scale and in the presence of threshold statistics, the mass fraction at PBH formation for small variances $(\sigma\ll \delta_{\rm th})$ reads as follows:
\bea \label{pbhmfrac}
\beta(M_{\rm PBH}) \simeq \frac{\gamma\sigma}{\sqrt{2\pi}\delta_{\rm th}}e^{-\frac{\delta^{2}_{\rm th}}{2\sigma^{2}}}~~~{\rm where}~~\sigma^{2} = \frac{16}{81}\int\;\frac{dk}{k}\;\left(\frac{k}{c_sk_s}\right)^{4}\;e^{-\frac{k^2}{2c^2_sk^2_s}}\;\Delta^{2}_{\zeta,{\bf RRR}}(k).
\eea
In figure ~(\ref{amplitudedistortion}) we have depicted the scalar power spectrum amplitude ($A$) necessary to achieve significant PBH abundance as a function of the transition wavenumber $k_{s}$ associated with the PBH mass, $M_{\rm PBH}/M_{\odot}$. The background involves constraints on the amplitude coming from the various experiments \cite{Planck:2018jri,Cyr:2023pgw,EPTA:2023xxk,bird2011minimally,Jeong:2014gna}. The magenta-coloured band, bounded by the brown and purple lines, highlights the amplitude region where PBH abundance lies within $f_{\rm PBH}\sim {\cal O}(10^{-3}-1)$. The obtained PBHs are very small, $M_{\rm PBH}\sim {\cal O}(10^{-33}-10^{-27})M_{\odot}$, and big, $M_{\rm PBH}\sim {\cal O}(10^{-6}-10^{-1})M_{\odot}$, within the PBHs mass fraction window, and they confront the most recent microlensing constraints. \textcolor{black}{Since in the figure ~(\ref{amplitudedistortion}) the observational windows from the various experiments \cite{Planck:2018jri,Cyr:2023pgw,EPTA:2023xxk,bird2011minimally,Jeong:2014gna} favoured the generation of the high-mass PBH formation within the range, $M_{\rm PBH}\sim {\cal O}(10^{-6}-10^{-1})M_{\odot}$ it is explicitly depicted here. However, for the general readers and curious authors we suggest to please look into the numerical analysis of PBH formation of ref. \cite{Choudhury:2024dei} for the complete understanding of the lower mass PBHs generation as well in the present context of discussion. For the better understanding purpose we have written table I where we have explicitly shown the generation of large as well as small mass PBHs which can overcome overproduction.}

\begin{table} \tiny
  \begin{tabular}{cccc|ccc}
    \toprule
    \multirow{2}{*}{Effective sound speed} &
      \multicolumn{3}{c|}{High mass} &
      \multicolumn{3}{c}{Low mass} \\
      $c_{s}$ & $M_{\rm PBH}/M_{\odot}$ & $f_{\rm PBH}$ & $k_{s}/{\rm Mpc^{-1}}$ & $M_{\rm PBH}/M_{\odot}$ & $f_{\rm PBH}$ & $k_{s}/{\rm Mpc^{-1}}$ \\
      \midrule\midrule
     & $(7.5\times 10^{-7}, 1.1 \times 10^{-6})$ & $(1,10^{-4})$ & $10^{8}$ & $(10^{-33.69},10^{-33.62})$ & $(1,10^{-4})$ & $10^{22}$ \\
    \cmidrule{2-7}
    1 & $(7.8\times 10^{-5},1\times 10^{-4})$ & $(0.018,10^{-4})$ & $10^{7}$ & $(10^{-31.71},10^{-31.64})$ & $(1,10^{-4})$ & $10^{21}$ \\
    \cmidrule{2-7}
     & $(7.7\times 10^{-3},9.2\times 10^{-3})$ & $(0.0028,10^{-4})$ & $10^{6}$ & $(10^{-29.74},10^{-29.66})$ & $(1,10^{-4})$ & $10^{20}$ \\
    \hline
     & $(5.1\times 10^{-6}, 3.4 \times 10^{-6})$ & $(0.246,10^{-4})$ & $8\times 10^{8}$ & $(10^{-33.23},10^{-33.15})$ & $(1,10^{-4})$ & $10^{22}$ \\
    \cmidrule{2-7}
    0.88 & $(3.6\times 10^{-5}, 2.4 \times 10^{-5})$ & $(0.035,10^{-4})$ & $10^{7}$ & $(10^{-31.25},10^{-31.17})$ & $(1,10^{-4})$ & $10^{21}$  \\
    \cmidrule{2-7}
     & $(4.4\times 10^{-4}, 2.8 \times 10^{-4})$ & $(0.006,10^{-4})$ & $3\times 10^{6}$ & $(10^{-29.28},10^{-29.19})$ & $(1,10^{-4})$ & $10^{20}$ \\
    \bottomrule
  \end{tabular}
  \label{tab1}
  \caption{ \textcolor{black}{An overview of the results obtained by different parameters utilizing the DRG-resummed scalar power spectrum, including PBH masses $M_{\rm PBH}$ in the high (near solar-mass) and low (very lower than solar-mass) mass regimes, their corresponding abundances $f_{\rm PBH}$, and the peak wavenumber $k_{s}$. The set of parameters are presented for two independent values of $c_{s}=1,0.88$.} }
\end{table}

\textcolor{black}{Nevertheless, the Press-Schechter formalism contains a number of restrictions, which are listed point-by-point below:}
\begin{enumerate}
    \item \textcolor{black}{In this formalism, the density contrast threshold value $\delta_{\rm th}$ has a significant impact on $f_{\rm PBH}$. However, it is evident from earlier numerical and analytical studies that the density contrast $\delta_{\rm th}$ threshold value for the formation of PBHs in the radiation-dominated era ranges from $0.3<\delta_{\rm th}<0.66$ \cite{Carr:1974nx,Mahbub:2019uhl,Harada:2013epa,Shibata:1999zs,Musco:2008hv,Polnarev:2006aa,Escriva:2019phb}. Most of the time, it is assumed that $\delta_{\rm th}=0.414$, which is exactly equivalent to the analytical conclusion found in ref. \cite{Harada:2013epa}.}

    \item \textcolor{black}{Additionally, the corresponding $f_{\rm PBH}$ is highly sensitive to the peak value of the power spectrum at $k=k_{\rm PBH}=k_{s}$, where the exact power amplification occurs. Consequently, a fine-tuning of the parameters of the underlying theory that describes the inflationary potential occurs to describe the production of PBHs with the necessarily required value of the PBHs abundance. This is because the variance $\sigma$ of the Gaussian probability distribution function depends upon the explicit expression for the primordial power spectrum. This specifies the essential criterion and does not apply to any particular model in this discussion if we take into account the mass fraction of monochromatic PBHs. See refs. \cite{Motohashi:2017kbs,Ballesteros:2017fsr} to know more about this issue.} 

    \item \textcolor{black}{It should be mentioned that additional ambiguity arises from selecting the right window function while calculating the variance. For the sake of simplicity, we have used a Gaussian window function in this context; however, alternative window function options can also be explored for the current purpose \cite{Young:2019osy,Ando:2018qdb}.}

    \item \textcolor{black}{There are more recent developments of this computation of the PBH fraction that suggest the use of more refined formalism, which is commonly known as peak theory, which can accommodate the non-Gaussian nature of the energy density fluctuations of the seed primordial perturbations $\delta$ due to their non-linear relation to the curvature perturbation variable $\zeta$ as explicitly discussed in the references \cite{Young:2019yug,DeLuca:2019qsy}. Using this fact in the references \cite{Choudhury:2024kjj,Choudhury:2024dzw}, we have further refined our computation of PBH fraction, incorporating the features of primordial non-Gaussianity within the framework of peak theory. We examine two distinct values of the non-Gaussianity parameter, $f_{\rm NL} = (-39.95, -35/8)$, found in the matter bounce and ekpyrotic contraction scenarios, respectively. We demonstrate that a negatively large amount of $f_{\rm NL}$ can provide a significant abundance, $10^{-3} \leq f_{\rm PBH} \leq 1$, and fully mitigate the problem of PBH overproduction. We further point out that the scalar-induced gravitational wave explanation of the most recent PTA (NANOGrav15 and EPTA) signal has an agreement under $1\sigma$ in the situation with the effective sound speed $c_s = 1$ paired with $f_{\rm NL} = -39.95$. This non-Gaussian analysis can further allow the PBH mass spectrum within the range $M_{\rm PBH}\sim 10^{-6}~M_{\odot}-0.1~M_{\odot}$, which was shown in references \cite{Choudhury:2024kjj,Choudhury:2024dzw}. It is important to note that, references \cite{Choudhury:2024kjj,Choudhury:2024dzw} considered the first-order NG correction is encoded by the parameter $f_{\rm NL}$, which represents the local form of NGs in the curvature perturbation. Additionally, we discovered in this study that the signature of this $f_{\rm NL}$ has a significant impact on the PBH abundance. In most circumstances, $f_{\rm NL} < 0$ is thought to be more favorable to suppress the PBH abundance, whereas positive NGs $f_{\rm NL} > 0$ invite the danger of overproduction.}

    \item \textcolor{black}{In addition to this, the critical collapse of PBH in terms of the computation of the threshold density $\delta_{\rm th}$ depends crucially on the shape of the overdensity region as discussed in the references \cite{Musco:2018rwt,Musco:2020jjb}. Also, this is not just a constant number as considered within the framework of Press-Schechter formalism. However, we have not restricted our analysis only up to this. In references \cite{Choudhury:2024kjj,Choudhury:2024dzw} we have discussed that with the help of peak theory once can explicitly study the critical collapse of PBHs by making use of the compaction function approach. We urge the readers to look into the refs. \cite{Choudhury:2024kjj,Choudhury:2024dzw} for more details on this issue.  }
\end{enumerate}

\section{Fine-tuning in USR phase and its resolution}

\textcolor{black}{One of the most efficient mechanisms to amplify the primordial curvature power spectrum on very small scales—and thereby produce an abundant population of primordial black holes (PBHs)—is an ultra slow-roll (USR) phase during the early universe. In standard slow-roll inflation the comoving curvature perturbation $\zeta$ is approximately constant after horizon crossing because the inflaton rolls smoothly down a sufficiently steep potential. In contrast, during USR the inflaton traverses an unusually flat region of the potential where the slope is nearly zero. Hubble friction then rapidly damps the inflaton velocity, invalidating the ordinary slow-roll approximation. A key consequence is that curvature perturbations continue to grow on super-horizon scales instead of freezing out, allowing the scalar power spectrum to increase by many orders of magnitude—from the CMB amplitude ${\cal O}(10^{-9})$ to values ${\cal O}(10^{-2})$ typically required for efficient PBH formation. This super-horizon growth makes USR a central ingredient of many inflationary constructions that aim to produce PBHs as dark matter candidates or seeds for early structure.}

\textcolor{black}{Despite its phenomenological appeal, the USR mechanism is often criticized for severe fine-tuning. See refs.\cite{Cole:2023wyx,Profumo:2026qpn,Profumo:2026kfy} for more details on this issue. The enhancement of the power spectrum is exponentially sensitive to both the duration of the USR epoch and the detailed shape of the inflationary potential near the flat region. Small shifts in the flatness, the position of a near-inflection point, or the inflaton’s initial velocity can change the power spectrum and hence the predicted PBH abundance by many orders of magnitude, because the PBH mass fraction depends exponentially on the fluctuation amplitude. Achieving a large but not excessive enhancement thus typically requires carefully adjusted potential features so the model produces the needed localized peak in the power spectrum while remaining consistent with CMB constraints on the spectral index, tensor-to-scalar ratio, and non-Gaussianity. The transition into and out of USR must also be tuned: too prolonged a USR can overproduce PBHs, generate large non-Gaussianities, or lead to eternal inflation; too brief a USR fails to amplify perturbations sufficiently. These narrow viable parameter ranges motivate alternative model-building efforts.}

\textcolor{black}{Several strategies have been proposed to relax the tuning in USR-based scenarios, including multi-field dynamics, non-canonical kinetic terms, controlled stochastic (quantum diffusion) effects, modified gravity, or repeated short USR episodes. Quantum diffusion can be particularly important in parts of parameter space, altering PBH abundance estimates and forcing a stochastic treatment of inflaton dynamics rather than a purely classical one. No single, universally accepted remedy has emerged, and the search continues for more robust mechanisms in which large small-scale enhancements arise naturally.}

\textcolor{black}{Non-singular bouncing cosmologies provide a qualitatively different route to PBH formation that can substantially mitigate fine-tuning issues. In these scenarios the universe undergoes a contracting phase before a nonsingular bounce to the expanding era, avoiding an initial Big Bang singularity. During a matter-dominated contraction (effective equation of state $w \simeq 0$), the comoving Hubble radius shrinks and modes continuously exit the horizon; crucially, the dominant curvature perturbation mode grows on super-horizon scales during contraction. This background-driven amplification increases density contrasts progressively, analogously to gravitational instability in a pressureless fluid, and can drive regions nonlinear and prone to collapse into PBHs without requiring localized peaks in the primordial spectrum at horizon crossing. Because the enhancement is driven by the global dynamics of contraction rather than delicate local features in a potential, PBH production in bouncing models can be far less sensitive to microscopic parameter choices.}

\textcolor{black}{Realistic bouncing scenarios must, however, address their own theoretical challenges: maintaining perturbative control during contraction, suppressing dangerous anisotropies associated with the BKL instability, ensuring a stable, nonsingular bounce free of ghosts or gradient instabilities, and transmitting the amplified perturbation spectrum into the expanding phase without excessive mode mixing. Model realizations often invoke ghost condensates, Horndeski/Galileon interactions, or quantum-gravity-inspired corrections to achieve a healthy bounce. The resulting PBH mass spectrum depends on the duration of matter contraction, the bounce energy scale, and the Hubble evolution, and can span a broad range with potential observational consequences—from dark matter contributions and seeds of supermassive black holes to stochastic gravitational-wave backgrounds and distinctive imprints on primordial spectra. Thus, while USR in inflation remains a compelling and flexible mechanism for PBH formation, non-singular bouncing cosmologies offer an attractive, dynamical alternative that can produce significant super-horizon enhancement with reduced fine-tuning, provided the model-specific stability and observational constraints are satisfied. Take a look at refs. \cite{Carr:2011hv,Quintin:2016qro,Chen:2016kjx,Clifton:2017hvg,Banerjee:2022xft,Chen:2022usd,Papanikolaou:2024fzf}, where PBH production in non-singular bounce situations was also investigated. In these cases, PBHs were created without fine-tuning because of the super-horizon increase of the curvature perturbations during a matter contracting phase.}

\section{Discussion}

This study constructs an EFT of bounce to investigate the formation of large-mass PBHs within single-field inflation, which is successfully evade the strong no-go theorem on PBH mass as proposed in \cite{Choudhury:2023vuj,Choudhury:2023jlt,Choudhury:2023rks}. The cosmological framework includes five phases: contraction, bounce, and the inflationary stages SRI, USR, and SRII. The scalar power spectrum is derived for each phase, and quantum loop corrections are analyzed using the Schwinger–Keldysh formalism. Divergences are treated through systematic regularization, renormalization, and Dynamical Renormalization Group resummation, yielding a finite spectrum including all loop effects. The resummed spectrum reaches amplitudes suitable for PBH formation while maintaining ~60 e-folds. Predictions for PBH masses, abundances, sound-speed effects, spectral distortions, and observational constraints are also examined.

\vskip 0.1in
\emph{Acknowledgments:}
SC would like to thank The North American Nanohertz Observatory for Gravitational Waves (NANOGrav) collaboration and the National Academy of Sciences (NASI), Prayagraj, India, for being elected as an associate member and the member of the academy respectively. SC acknowledge our debt to the people
belonging to various parts of the world for their generous and steady support for research in natural sciences.


\renewcommand{\leftmark}{\MakeUppercase{Bibliography}}
\phantomsection
\bibliography{Refs}

\providecommand{\href}[2]{#2}\begingroup\raggedright\begin{thebibliography}{100}

\bibitem{Zeldovich:1967lct}
Y.~B. Zel'dovich and I.~D. Novikov, ``{The Hypothesis of Cores Retarded during
  Expansion and the Hot Cosmological Model},'' {\em Soviet Astron. AJ (Engl.
  Transl. ),} {\bfseries 10} (1967) 602.

\bibitem{Hawking:1974rv}
S.~W. Hawking, ``{Black hole explosions},''
  \href{http://dx.doi.org/10.1038/248030a0}{{\em Nature} {\bfseries 248} (1974)
  30--31}.

\bibitem{Carr:1974nx}
B.~J. Carr and S.~W. Hawking, ``{Black holes in the early Universe},''
  \href{http://dx.doi.org/10.1093/mnras/168.2.399}{{\em Mon. Not. Roy. Astron.
  Soc.} {\bfseries 168} (1974) 399--415}.

\bibitem{Carr:1975qj}
B.~J. Carr, ``{The Primordial black hole mass spectrum},''
  \href{http://dx.doi.org/10.1086/153853}{{\em Astrophys. J.} {\bfseries 201}
  (1975) 1--19}.

\bibitem{Chapline:1975ojl}
G.~F. Chapline, ``{Cosmological effects of primordial black holes},''
  \href{http://dx.doi.org/10.1038/253251a0}{{\em Nature} {\bfseries 253}
  no.~5489, (1975) 251--252}.

\bibitem{Carr:1993aq}
B.~J. Carr and J.~E. Lidsey, ``{Primordial black holes and generalized
  constraints on chaotic inflation},''
  \href{http://dx.doi.org/10.1103/PhysRevD.48.543}{{\em Phys. Rev. D}
  {\bfseries 48} (1993) 543--553}.

\bibitem{Choudhury:2011jt}
S.~Choudhury and S.~Pal, ``{Fourth level MSSM inflation from new flat
  directions},'' \href{http://dx.doi.org/10.1088/1475-7516/2012/04/018}{{\em
  JCAP} {\bfseries 04} (2012) 018},
  \href{http://arxiv.org/abs/1111.3441}{{\ttfamily arXiv:1111.3441 [hep-ph]}}.

\bibitem{Yokoyama:1998pt}
J.~Yokoyama, ``{Chaotic new inflation and formation of primordial black
  holes},'' \href{http://dx.doi.org/10.1103/PhysRevD.58.083510}{{\em Phys. Rev.
  D} {\bfseries 58} (1998) 083510},
  \href{http://arxiv.org/abs/astro-ph/9802357}{{\ttfamily
  arXiv:astro-ph/9802357}}.

\bibitem{Kawasaki:1998vx}
M.~Kawasaki and T.~Yanagida, ``{Primordial black hole formation in
  supergravity},'' \href{http://dx.doi.org/10.1103/PhysRevD.59.043512}{{\em
  Phys. Rev. D} {\bfseries 59} (1999) 043512},
  \href{http://arxiv.org/abs/hep-ph/9807544}{{\ttfamily arXiv:hep-ph/9807544}}.

\bibitem{Rubin:2001yw}
S.~G. Rubin, A.~S. Sakharov, and M.~Y. Khlopov, ``{The Formation of primary
  galactic nuclei during phase transitions in the early universe},''
  \href{http://dx.doi.org/10.1134/1.1385631}{{\em J. Exp. Theor. Phys.}
  {\bfseries 91} (2001) 921--929},
  \href{http://arxiv.org/abs/hep-ph/0106187}{{\ttfamily arXiv:hep-ph/0106187}}.

\bibitem{Khlopov:2002yi}
M.~Y. Khlopov, S.~G. Rubin, and A.~S. Sakharov, ``{Strong primordial
  inhomogeneities and galaxy formation},''
  \href{http://arxiv.org/abs/astro-ph/0202505}{{\ttfamily
  arXiv:astro-ph/0202505}}.

\bibitem{Khlopov:2004sc}
M.~Y. Khlopov, S.~G. Rubin, and A.~S. Sakharov, ``{Primordial structure of
  massive black hole clusters},''
  \href{http://dx.doi.org/10.1016/j.astropartphys.2004.12.002}{{\em Astropart.
  Phys.} {\bfseries 23} (2005) 265},
  \href{http://arxiv.org/abs/astro-ph/0401532}{{\ttfamily
  arXiv:astro-ph/0401532}}.

\bibitem{Saito:2008em}
R.~Saito, J.~Yokoyama, and R.~Nagata, ``{Single-field inflation, anomalous
  enhancement of superhorizon fluctuations, and non-Gaussianity in primordial
  black hole formation},''
  \href{http://dx.doi.org/10.1088/1475-7516/2008/06/024}{{\em JCAP} {\bfseries
  06} (2008) 024}, \href{http://arxiv.org/abs/0804.3470}{{\ttfamily
  arXiv:0804.3470 [astro-ph]}}.

\bibitem{Khlopov:2008qy}
M.~Y. Khlopov, ``{Primordial Black Holes},''
  \href{http://dx.doi.org/10.1088/1674-4527/10/6/001}{{\em Res. Astron.
  Astrophys.} {\bfseries 10} (2010) 495--528},
  \href{http://arxiv.org/abs/0801.0116}{{\ttfamily arXiv:0801.0116
  [astro-ph]}}.

\bibitem{Carr:2009jm}
B.~J. Carr, K.~Kohri, Y.~Sendouda, and J.~Yokoyama, ``{New cosmological
  constraints on primordial black holes},''
  \href{http://dx.doi.org/10.1103/PhysRevD.81.104019}{{\em Phys. Rev. D}
  {\bfseries 81} (2010) 104019},
  \href{http://arxiv.org/abs/0912.5297}{{\ttfamily arXiv:0912.5297
  [astro-ph.CO]}}.

\bibitem{Lyth:2011kj}
D.~H. Lyth, ``{Primordial black hole formation and hybrid inflation},''
  \href{http://arxiv.org/abs/1107.1681}{{\ttfamily arXiv:1107.1681
  [astro-ph.CO]}}.

\bibitem{Drees:2011yz}
M.~Drees and E.~Erfani, ``{Running Spectral Index and Formation of Primordial
  Black Hole in Single Field Inflation Models},''
  \href{http://dx.doi.org/10.1088/1475-7516/2012/01/035}{{\em JCAP} {\bfseries
  01} (2012) 035}, \href{http://arxiv.org/abs/1110.6052}{{\ttfamily
  arXiv:1110.6052 [astro-ph.CO]}}.

\bibitem{Drees:2011hb}
M.~Drees and E.~Erfani, ``{Running-Mass Inflation Model and Primordial Black
  Holes},'' \href{http://dx.doi.org/10.1088/1475-7516/2011/04/005}{{\em JCAP}
  {\bfseries 04} (2011) 005}, \href{http://arxiv.org/abs/1102.2340}{{\ttfamily
  arXiv:1102.2340 [hep-ph]}}.

\bibitem{Ezquiaga:2017fvi}
J.~M. Ezquiaga, J.~Garcia-Bellido, and E.~Ruiz~Morales, ``{Primordial Black
  Hole production in Critical Higgs Inflation},''
  \href{http://dx.doi.org/10.1016/j.physletb.2017.11.039}{{\em Phys. Lett. B}
  {\bfseries 776} (2018) 345--349},
  \href{http://arxiv.org/abs/1705.04861}{{\ttfamily arXiv:1705.04861
  [astro-ph.CO]}}.

\bibitem{Kannike:2017bxn}
K.~Kannike, L.~Marzola, M.~Raidal, and H.~Veerm\"ae, ``{Single Field Double
  Inflation and Primordial Black Holes},''
  \href{http://dx.doi.org/10.1088/1475-7516/2017/09/020}{{\em JCAP} {\bfseries
  09} (2017) 020}, \href{http://arxiv.org/abs/1705.06225}{{\ttfamily
  arXiv:1705.06225 [astro-ph.CO]}}.

\bibitem{Hertzberg:2017dkh}
M.~P. Hertzberg and M.~Yamada, ``{Primordial Black Holes from Polynomial
  Potentials in Single Field Inflation},''
  \href{http://dx.doi.org/10.1103/PhysRevD.97.083509}{{\em Phys. Rev. D}
  {\bfseries 97} no.~8, (2018) 083509},
  \href{http://arxiv.org/abs/1712.09750}{{\ttfamily arXiv:1712.09750
  [astro-ph.CO]}}.

\bibitem{Pi:2017gih}
S.~Pi, Y.-l. Zhang, Q.-G. Huang, and M.~Sasaki, ``{Scalaron from $R^2$-gravity
  as a heavy field},''
  \href{http://dx.doi.org/10.1088/1475-7516/2018/05/042}{{\em JCAP} {\bfseries
  05} (2018) 042}, \href{http://arxiv.org/abs/1712.09896}{{\ttfamily
  arXiv:1712.09896 [astro-ph.CO]}}.

\bibitem{Gao:2018pvq}
T.-J. Gao and Z.-K. Guo, ``{Primordial Black Hole Production in Inflationary
  Models of Supergravity with a Single Chiral Superfield},''
  \href{http://dx.doi.org/10.1103/PhysRevD.98.063526}{{\em Phys. Rev. D}
  {\bfseries 98} no.~6, (2018) 063526},
  \href{http://arxiv.org/abs/1806.09320}{{\ttfamily arXiv:1806.09320
  [hep-ph]}}.

\bibitem{Dalianis:2018frf}
I.~Dalianis, A.~Kehagias, and G.~Tringas, ``{Primordial black holes from
  \ensuremath{\alpha}-attractors},''
  \href{http://dx.doi.org/10.1088/1475-7516/2019/01/037}{{\em JCAP} {\bfseries
  01} (2019) 037}, \href{http://arxiv.org/abs/1805.09483}{{\ttfamily
  arXiv:1805.09483 [astro-ph.CO]}}.

\bibitem{Cicoli:2018asa}
M.~Cicoli, V.~A. Diaz, and F.~G. Pedro, ``{Primordial Black Holes from String
  Inflation},'' \href{http://dx.doi.org/10.1088/1475-7516/2018/06/034}{{\em
  JCAP} {\bfseries 06} (2018) 034},
  \href{http://arxiv.org/abs/1803.02837}{{\ttfamily arXiv:1803.02837
  [hep-th]}}.

\bibitem{Choudhury:2013woa}
S.~Choudhury and A.~Mazumdar, ``{Primordial blackholes and gravitational waves
  for an inflection-point model of inflation},''
  \href{http://dx.doi.org/10.1016/j.physletb.2014.04.050}{{\em Phys. Lett. B}
  {\bfseries 733} (2014) 270--275},
  \href{http://arxiv.org/abs/1307.5119}{{\ttfamily arXiv:1307.5119
  [astro-ph.CO]}}.

\bibitem{Sasaki:2016jop}
M.~Sasaki, T.~Suyama, T.~Tanaka, and S.~Yokoyama, ``{Primordial Black Hole
  Scenario for the Gravitational-Wave Event GW150914},''
  \href{http://dx.doi.org/10.1103/PhysRevLett.117.061101}{{\em Phys. Rev.
  Lett.} {\bfseries 117} no.~6, (2016) 061101},
  \href{http://arxiv.org/abs/1603.08338}{{\ttfamily arXiv:1603.08338
  [astro-ph.CO]}}. [Erratum: Phys.Rev.Lett. 121, 059901 (2018)].

\bibitem{Raidal:2017mfl}
M.~Raidal, V.~Vaskonen, and H.~Veerm\"ae, ``{Gravitational Waves from
  Primordial Black Hole Mergers},''
  \href{http://dx.doi.org/10.1088/1475-7516/2017/09/037}{{\em JCAP} {\bfseries
  09} (2017) 037}, \href{http://arxiv.org/abs/1707.01480}{{\ttfamily
  arXiv:1707.01480 [astro-ph.CO]}}.

\bibitem{Papanikolaou:2020qtd}
T.~Papanikolaou, V.~Vennin, and D.~Langlois, ``{Gravitational waves from a
  universe filled with primordial black holes},''
  \href{http://dx.doi.org/10.1088/1475-7516/2021/03/053}{{\em JCAP} {\bfseries
  03} (2021) 053}, \href{http://arxiv.org/abs/2010.11573}{{\ttfamily
  arXiv:2010.11573 [astro-ph.CO]}}.

\bibitem{Ashoorioon:2022raz}
A.~Ashoorioon, K.~Rezazadeh, and A.~Rostami, ``{NANOGrav signal from the end of
  inflation and the LIGO mass and heavier primordial black holes},''
  \href{http://dx.doi.org/10.1016/j.physletb.2022.137542}{{\em Phys. Lett. B}
  {\bfseries 835} (2022) 137542},
  \href{http://arxiv.org/abs/2202.01131}{{\ttfamily arXiv:2202.01131
  [astro-ph.CO]}}.

\bibitem{Papanikolaou:2022chm}
T.~Papanikolaou, ``{Gravitational waves induced from primordial black hole
  fluctuations: the~effect of an extended mass function},''
  \href{http://dx.doi.org/10.1088/1475-7516/2022/10/089}{{\em JCAP} {\bfseries
  10} (2022) 089}, \href{http://arxiv.org/abs/2207.11041}{{\ttfamily
  arXiv:2207.11041 [astro-ph.CO]}}.

\bibitem{Papanikolaou:2023crz}
T.~Papanikolaou, ``{Primordial black holes in loop quantum cosmology: the
  effect on the threshold},''
  \href{http://dx.doi.org/10.1088/1361-6382/acd97d}{{\em Class. Quant. Grav.}
  {\bfseries 40} no.~13, (2023) 134001},
  \href{http://arxiv.org/abs/2301.11439}{{\ttfamily arXiv:2301.11439 [gr-qc]}}.

\bibitem{Wang:2022nml}
X.~Wang, Y.-l. Zhang, R.~Kimura, and M.~Yamaguchi, ``{Reconstruction of Power
  Spectrum of Primordial Curvature Perturbations on small scales from
  Primordial Black Hole Binaries scenario of LIGO/VIRGO detection},''
  \href{http://arxiv.org/abs/2209.12911}{{\ttfamily arXiv:2209.12911
  [astro-ph.CO]}}.

\bibitem{Riotto:2023hoz}
A.~Riotto, ``{The Primordial Black Hole Formation from Single-Field Inflation
  is Not Ruled Out},'' \href{http://arxiv.org/abs/2301.00599}{{\ttfamily
  arXiv:2301.00599 [astro-ph.CO]}}.

\bibitem{Riotto:2023gpm}
A.~Riotto, ``{The Primordial Black Hole Formation from Single-Field Inflation
  is Still Not Ruled Out},'' \href{http://arxiv.org/abs/2303.01727}{{\ttfamily
  arXiv:2303.01727 [astro-ph.CO]}}.

\bibitem{Papanikolaou:2022did}
T.~Papanikolaou, A.~Lymperis, S.~Lola, and E.~N. Saridakis, ``{Primordial black
  holes and gravitational waves from non-canonical inflation},''
  \href{http://dx.doi.org/10.1088/1475-7516/2023/03/003}{{\em JCAP} {\bfseries
  03} (2023) 003}, \href{http://arxiv.org/abs/2211.14900}{{\ttfamily
  arXiv:2211.14900 [astro-ph.CO]}}.

\bibitem{Choudhury:2023vuj}
S.~Choudhury, M.~R. Gangopadhyay, and M.~Sami, ``{No-go for the formation of
  heavy mass Primordial Black Holes in Single Field Inflation},''
  \href{http://dx.doi.org/10.1140/epjc/s10052-024-13218-2}{{\em Eur. Phys. J.
  C} {\bfseries 84} no.~9, (2024) 884},
  \href{http://arxiv.org/abs/2301.10000}{{\ttfamily arXiv:2301.10000
  [astro-ph.CO]}}.

\bibitem{Choudhury:2023jlt}
S.~Choudhury, S.~Panda, and M.~Sami, ``{PBH formation in EFT of single field
  inflation with sharp transition},''
  \href{http://dx.doi.org/10.1016/j.physletb.2023.138123}{{\em Phys. Lett. B}
  {\bfseries 845} (2023) 138123},
  \href{http://arxiv.org/abs/2302.05655}{{\ttfamily arXiv:2302.05655
  [astro-ph.CO]}}.

\bibitem{Choudhury:2023rks}
S.~Choudhury, S.~Panda, and M.~Sami, ``{Quantum loop effects on the power
  spectrum and constraints on primordial black holes},''
  \href{http://dx.doi.org/10.1088/1475-7516/2023/11/066}{{\em JCAP} {\bfseries
  11} (2023) 066}, \href{http://arxiv.org/abs/2303.06066}{{\ttfamily
  arXiv:2303.06066 [astro-ph.CO]}}.

\bibitem{Choudhury:2023hvf}
S.~Choudhury, S.~Panda, and M.~Sami, ``{Galileon inflation evades the no-go for
  PBH formation in the single-field framework},''
  \href{http://dx.doi.org/10.1088/1475-7516/2023/08/078}{{\em JCAP} {\bfseries
  08} (2023) 078}, \href{http://arxiv.org/abs/2304.04065}{{\ttfamily
  arXiv:2304.04065 [astro-ph.CO]}}.

\bibitem{Choudhury:2023kdb}
S.~Choudhury, A.~Karde, S.~Panda, and M.~Sami, ``{Primordial non-Gaussianity
  from ultra slow-roll Galileon inflation},''
  \href{http://dx.doi.org/10.1088/1475-7516/2024/01/012}{{\em JCAP} {\bfseries
  01} (2024) 012}, \href{http://arxiv.org/abs/2306.12334}{{\ttfamily
  arXiv:2306.12334 [astro-ph.CO]}}.

\bibitem{Choudhury:2023hfm}
S.~Choudhury, A.~Karde, S.~Panda, and M.~Sami, ``{Scalar induced gravity waves
  from ultra slow-roll galileon inflation},''
  \href{http://dx.doi.org/10.1016/j.nuclphysb.2024.116678}{{\em Nucl. Phys. B}
  {\bfseries 1007} (2024) 116678},
  \href{http://arxiv.org/abs/2308.09273}{{\ttfamily arXiv:2308.09273
  [astro-ph.CO]}}.

\bibitem{Bhattacharya:2023ysp}
G.~Bhattacharya, S.~Choudhury, K.~Dey, S.~Ghosh, A.~Karde, and N.~S. Mishra,
  ``{Evading no-go for PBH formation and production of SIGWs using Multiple
  Sharp Transitions in EFT of single field inflation},''
  \href{http://dx.doi.org/10.1016/j.dark.2024.101602}{{\em Phys. Dark Univ.}
  {\bfseries 46} (2024) 101602},
  \href{http://arxiv.org/abs/2309.00973}{{\ttfamily arXiv:2309.00973
  [astro-ph.CO]}}.

\bibitem{Choudhury:2023fwk}
S.~Choudhury, K.~Dey, A.~Karde, S.~Panda, and M.~Sami, ``{Primordial
  non-Gaussianity as a saviour for PBH overproduction in SIGWs generated by
  pulsar timing arrays for Galileon inflation},''
  \href{http://dx.doi.org/10.1016/j.physletb.2024.138925}{{\em Phys. Lett. B}
  {\bfseries 856} (2024) 138925},
  \href{http://arxiv.org/abs/2310.11034}{{\ttfamily arXiv:2310.11034
  [astro-ph.CO]}}.

\bibitem{Choudhury:2023fjs}
S.~Choudhury, K.~Dey, and A.~Karde, ``{Untangling PBH overproduction in
  $w$-SIGWs generated by Pulsar Timing Arrays for MST-EFT of single field
  inflation},'' \href{http://arxiv.org/abs/2311.15065}{{\ttfamily
  arXiv:2311.15065 [astro-ph.CO]}}.

\bibitem{Choudhury:2024one}
S.~Choudhury, A.~Karde, S.~Panda, and M.~Sami, ``{Realisation of the ultra-slow
  roll phase in Galileon inflation and PBH overproduction},''
  \href{http://dx.doi.org/10.1088/1475-7516/2024/07/034}{{\em JCAP} {\bfseries
  07} (2024) 034}, \href{http://arxiv.org/abs/2401.10925}{{\ttfamily
  arXiv:2401.10925 [astro-ph.CO]}}.

\bibitem{Choudhury:2024ybk}
S.~Choudhury, ``{Large fluctuations in the sky},''
  \href{http://dx.doi.org/10.1142/S0218271824410074}{{\em Int. J. Mod. Phys. D}
  {\bfseries 33} no.~15, (2024) 2441007},
  \href{http://arxiv.org/abs/2403.07343}{{\ttfamily arXiv:2403.07343
  [astro-ph.CO]}}.

\bibitem{Choudhury:2024jlz}
S.~Choudhury, A.~Karde, P.~Padiyar, and M.~Sami, ``{Primordial Black Holes from
  Effective Field Theory of Stochastic Single Field Inflation at NNNLO},''
  \href{http://arxiv.org/abs/2403.13484}{{\ttfamily arXiv:2403.13484
  [astro-ph.CO]}}.

\bibitem{Firouzjahi:2023ahg}
H.~Firouzjahi and A.~Riotto, ``{Primordial Black Holes and loops in
  single-field inflation},''
  \href{http://dx.doi.org/10.1088/1475-7516/2024/02/021}{{\em JCAP} {\bfseries
  02} (2024) 021}, \href{http://arxiv.org/abs/2304.07801}{{\ttfamily
  arXiv:2304.07801 [astro-ph.CO]}}.

\bibitem{Firouzjahi:2023aum}
H.~Firouzjahi, ``{One-loop corrections in power spectrum in single field
  inflation},'' \href{http://dx.doi.org/10.1088/1475-7516/2023/10/006}{{\em
  JCAP} {\bfseries 10} (2023) 006},
  \href{http://arxiv.org/abs/2303.12025}{{\ttfamily arXiv:2303.12025
  [astro-ph.CO]}}.

\bibitem{Iacconi:2023ggt}
L.~Iacconi, D.~Mulryne, and D.~Seery, ``{Loop corrections in the separate
  universe picture},''
  \href{http://dx.doi.org/10.1088/1475-7516/2024/06/062}{{\em JCAP} {\bfseries
  06} (2024) 062}, \href{http://arxiv.org/abs/2312.12424}{{\ttfamily
  arXiv:2312.12424 [astro-ph.CO]}}.

\bibitem{Davies:2023hhn}
M.~W. Davies, L.~Iacconi, and D.~J. Mulryne, ``{Numerical 1-loop correction
  from a potential yielding ultra-slow-roll dynamics},''
  \href{http://dx.doi.org/10.1088/1475-7516/2024/04/050}{{\em JCAP} {\bfseries
  04} (2024) 050}, \href{http://arxiv.org/abs/2312.05694}{{\ttfamily
  arXiv:2312.05694 [astro-ph.CO]}}.

\bibitem{Jackson:2023obv}
J.~H.~P. Jackson, H.~Assadullahi, A.~D. Gow, K.~Koyama, V.~Vennin, and
  D.~Wands, ``{The separate-universe approach and sudden transitions during
  inflation},'' \href{http://dx.doi.org/10.1088/1475-7516/2024/05/053}{{\em
  JCAP} {\bfseries 05} (2024) 053},
  \href{http://arxiv.org/abs/2311.03281}{{\ttfamily arXiv:2311.03281
  [astro-ph.CO]}}.

\bibitem{Riotto:2024ayo}
A.~Riotto and J.~Silk, ``{The Future of Primordial Black Holes: Open Questions
  and Roadmap},'' \href{http://arxiv.org/abs/2403.02907}{{\ttfamily
  arXiv:2403.02907 [astro-ph.CO]}}.

\bibitem{Banerjee:2021lqu}
S.~Banerjee, S.~Choudhury, S.~Chowdhury, J.~Knaute, S.~Panda, and K.~Shirish,
  ``{Thermalization in quenched open quantum cosmology},''
  \href{http://dx.doi.org/10.1016/j.nuclphysb.2023.116368}{{\em Nucl. Phys. B}
  {\bfseries 996} (2023) 116368},
  \href{http://arxiv.org/abs/2104.10692}{{\ttfamily arXiv:2104.10692
  [hep-th]}}.

\bibitem{Choudhury:2023kam}
S.~Choudhury, ``{Single field inflation in the light of Pulsar Timing Array
  Data: quintessential interpretation of blue tilted tensor spectrum through
  Non-Bunch Davies initial condition},''
  \href{http://dx.doi.org/10.1140/epjc/s10052-024-12625-9}{{\em Eur. Phys. J.
  C} {\bfseries 84} no.~3, (2024) 278},
  \href{http://arxiv.org/abs/2307.03249}{{\ttfamily arXiv:2307.03249
  [astro-ph.CO]}}.

\bibitem{Choudhury:2024dei}
S.~Choudhury, A.~Karde, S.~Panda, and S.~SenGupta,
  ``{Regularized-renormalized-resummed loop corrected power spectrum of
  non-singular bounce with Primordial Black Hole formation},''
  \href{http://dx.doi.org/10.1140/epjc/s10052-024-13460-8}{{\em Eur. Phys. J.
  C} {\bfseries 84} no.~11, (2024) 1149},
  \href{http://arxiv.org/abs/2405.06882}{{\ttfamily arXiv:2405.06882
  [astro-ph.CO]}}.

\bibitem{Choudhury:2024dzw}
S.~Choudhury, S.~Ganguly, S.~Panda, S.~SenGupta, and P.~Tiwari, ``{Obviating
  PBH overproduction for SIGWs generated by pulsar timing arrays in loop
  corrected EFT of bounce},''
  \href{http://dx.doi.org/10.1088/1475-7516/2024/09/013}{{\em JCAP} {\bfseries
  09} (2024) 013}, \href{http://arxiv.org/abs/2407.18976}{{\ttfamily
  arXiv:2407.18976 [astro-ph.CO]}}.

\bibitem{Choudhury:2024aji}
S.~Choudhury and M.~Sami, ``{Large fluctuations and Primordial Black Holes},''
  \href{http://arxiv.org/abs/2407.17006}{{\ttfamily arXiv:2407.17006 [gr-qc]}}.

\bibitem{Choudhury:2024kjj}
S.~Choudhury, K.~Dey, S.~Ganguly, A.~Karde, S.~K. Singh, and P.~Tiwari,
  ``{Negative non-Gaussianity as a salvager for PBHs with PTAs in bounce},''
  \href{http://dx.doi.org/10.1140/epjc/s10052-025-14176-z}{{\em Eur. Phys. J.
  C} {\bfseries 85} no.~4, (2025) 472},
  \href{http://arxiv.org/abs/2409.18983}{{\ttfamily arXiv:2409.18983
  [astro-ph.CO]}}.

\bibitem{Choudhury:2024ezx}
S.~Choudhury, ``{Reconstructing inflationary potential from NANOGrav 15-year
  data: A robust study using Non-Bunch Davies initial condition},''
  \href{http://dx.doi.org/10.1016/j.jheap.2024.10.003}{{\em JHEAp} {\bfseries
  44} (2024) 220--242}, \href{http://arxiv.org/abs/2410.11893}{{\ttfamily
  arXiv:2410.11893 [gr-qc]}}.

\bibitem{Choudhury:2025kxg}
S.~Choudhury, ``{Stochastic origin of primordial fluctuations in the sky},''
  \href{http://dx.doi.org/10.1142/S0218271825440237}{{\em Int. J. Mod. Phys. D}
  {\bfseries 34} no.~16, (2025) 2544023},
  \href{http://arxiv.org/abs/2503.17635}{{\ttfamily arXiv:2503.17635 [gr-qc]}}.

\bibitem{Kristiano:2022maq}
J.~Kristiano and J.~Yokoyama, ``{Constraining Primordial Black Hole Formation
  from Single-Field Inflation},''
  \href{http://dx.doi.org/10.1103/PhysRevLett.132.221003}{{\em Phys. Rev.
  Lett.} {\bfseries 132} no.~22, (2024) 221003},
  \href{http://arxiv.org/abs/2211.03395}{{\ttfamily arXiv:2211.03395
  [hep-th]}}.

\bibitem{Kristiano:2023scm}
J.~Kristiano and J.~Yokoyama, ``{Note on the bispectrum and one-loop
  corrections in single-field inflation with primordial black hole
  formation},'' \href{http://dx.doi.org/10.1103/PhysRevD.109.103541}{{\em Phys.
  Rev. D} {\bfseries 109} no.~10, (2024) 103541},
  \href{http://arxiv.org/abs/2303.00341}{{\ttfamily arXiv:2303.00341
  [hep-th]}}.

\bibitem{Firouzjahi:2023bkt}
H.~Firouzjahi, ``{Revisiting loop corrections in single field ultraslow-roll
  inflation},'' \href{http://dx.doi.org/10.1103/PhysRevD.109.043514}{{\em Phys.
  Rev. D} {\bfseries 109} no.~4, (2024) 043514},
  \href{http://arxiv.org/abs/2311.04080}{{\ttfamily arXiv:2311.04080
  [astro-ph.CO]}}.

\bibitem{Khoury:2001wf}
J.~Khoury, B.~A. Ovrut, P.~J. Steinhardt, and N.~Turok, ``{The Ekpyrotic
  universe: Colliding branes and the origin of the hot big bang},''
  \href{http://dx.doi.org/10.1103/PhysRevD.64.123522}{{\em Phys. Rev. D}
  {\bfseries 64} (2001) 123522},
  \href{http://arxiv.org/abs/hep-th/0103239}{{\ttfamily arXiv:hep-th/0103239}}.

\bibitem{Khoury:2001zk}
J.~Khoury, B.~A. Ovrut, P.~J. Steinhardt, and N.~Turok, ``{Density
  perturbations in the ekpyrotic scenario},''
  \href{http://dx.doi.org/10.1103/PhysRevD.66.046005}{{\em Phys. Rev. D}
  {\bfseries 66} (2002) 046005},
  \href{http://arxiv.org/abs/hep-th/0109050}{{\ttfamily arXiv:hep-th/0109050}}.

\bibitem{Khoury:2001bz}
J.~Khoury, B.~A. Ovrut, N.~Seiberg, P.~J. Steinhardt, and N.~Turok, ``{From big
  crunch to big bang},''
  \href{http://dx.doi.org/10.1103/PhysRevD.65.086007}{{\em Phys. Rev. D}
  {\bfseries 65} (2002) 086007},
  \href{http://arxiv.org/abs/hep-th/0108187}{{\ttfamily arXiv:hep-th/0108187}}.

\bibitem{Buchbinder:2007ad}
E.~I. Buchbinder, J.~Khoury, and B.~A. Ovrut, ``{New Ekpyrotic cosmology},''
  \href{http://dx.doi.org/10.1103/PhysRevD.76.123503}{{\em Phys. Rev. D}
  {\bfseries 76} (2007) 123503},
  \href{http://arxiv.org/abs/hep-th/0702154}{{\ttfamily arXiv:hep-th/0702154}}.

\bibitem{Lehners:2007ac}
J.-L. Lehners, P.~McFadden, N.~Turok, and P.~J. Steinhardt, ``{Generating
  ekpyrotic curvature perturbations before the big bang},''
  \href{http://dx.doi.org/10.1103/PhysRevD.76.103501}{{\em Phys. Rev. D}
  {\bfseries 76} (2007) 103501},
  \href{http://arxiv.org/abs/hep-th/0702153}{{\ttfamily arXiv:hep-th/0702153}}.

\bibitem{Lehners:2008vx}
J.-L. Lehners, ``{Ekpyrotic and Cyclic Cosmology},''
  \href{http://dx.doi.org/10.1016/j.physrep.2008.06.001}{{\em Phys. Rept.}
  {\bfseries 465} (2008) 223--263},
  \href{http://arxiv.org/abs/0806.1245}{{\ttfamily arXiv:0806.1245
  [astro-ph]}}.

\bibitem{Brandenberger:2012zb}
R.~H. Brandenberger, ``{The Matter Bounce Alternative to Inflationary
  Cosmology},'' \href{http://arxiv.org/abs/1206.4196}{{\ttfamily
  arXiv:1206.4196 [astro-ph.CO]}}.

\bibitem{Chowdhury:2015cma}
D.~Chowdhury, V.~Sreenath, and L.~Sriramkumar, ``{The tensor bi-spectrum in a
  matter bounce},'' \href{http://dx.doi.org/10.1088/1475-7516/2015/11/002}{{\em
  JCAP} {\bfseries 11} (2015) 002},
  \href{http://arxiv.org/abs/1506.06475}{{\ttfamily arXiv:1506.06475
  [astro-ph.CO]}}.

\bibitem{Cai:2011tc}
Y.-F. Cai, S.-H. Chen, J.~B. Dent, S.~Dutta, and E.~N. Saridakis, ``{Matter
  Bounce Cosmology with the f(T) Gravity},''
  \href{http://dx.doi.org/10.1088/0264-9381/28/21/215011}{{\em Class. Quant.
  Grav.} {\bfseries 28} (2011) 215011},
  \href{http://arxiv.org/abs/1104.4349}{{\ttfamily arXiv:1104.4349
  [astro-ph.CO]}}.

\bibitem{Brandenberger:2016vhg}
R.~Brandenberger and P.~Peter, ``{Bouncing Cosmologies: Progress and
  Problems},'' \href{http://dx.doi.org/10.1007/s10701-016-0057-0}{{\em Found.
  Phys.} {\bfseries 47} no.~6, (2017) 797--850},
  \href{http://arxiv.org/abs/1603.05834}{{\ttfamily arXiv:1603.05834
  [hep-th]}}.

\bibitem{Boyle:2004gv}
L.~A. Boyle, P.~J. Steinhardt, and N.~Turok, ``{A New duality relating density
  perturbations in expanding and contracting Friedmann cosmologies},''
  \href{http://dx.doi.org/10.1103/PhysRevD.70.023504}{{\em Phys. Rev. D}
  {\bfseries 70} (2004) 023504},
  \href{http://arxiv.org/abs/hep-th/0403026}{{\ttfamily arXiv:hep-th/0403026}}.

\bibitem{Wands:1998yp}
D.~Wands, ``{Duality invariance of cosmological perturbation spectra},''
  \href{http://dx.doi.org/10.1103/PhysRevD.60.023507}{{\em Phys. Rev. D}
  {\bfseries 60} (1999) 023507},
  \href{http://arxiv.org/abs/gr-qc/9809062}{{\ttfamily arXiv:gr-qc/9809062}}.

\bibitem{Peter:2002cn}
P.~Peter and N.~Pinto-Neto, ``{Primordial perturbations in a non singular
  bouncing universe model},''
  \href{http://dx.doi.org/10.1103/PhysRevD.66.063509}{{\em Phys. Rev. D}
  {\bfseries 66} (2002) 063509},
  \href{http://arxiv.org/abs/hep-th/0203013}{{\ttfamily arXiv:hep-th/0203013}}.

\bibitem{Allen:2004vz}
L.~E. Allen and D.~Wands, ``{Cosmological perturbations through a simple
  bounce},'' \href{http://dx.doi.org/10.1103/PhysRevD.70.063515}{{\em Phys.
  Rev. D} {\bfseries 70} (2004) 063515},
  \href{http://arxiv.org/abs/astro-ph/0404441}{{\ttfamily
  arXiv:astro-ph/0404441}}.

\bibitem{Martin:2003sf}
J.~Martin and P.~Peter, ``{Parametric amplification of metric fluctuations
  through a bouncing phase},''
  \href{http://dx.doi.org/10.1103/PhysRevD.68.103517}{{\em Phys. Rev. D}
  {\bfseries 68} (2003) 103517},
  \href{http://arxiv.org/abs/hep-th/0307077}{{\ttfamily arXiv:hep-th/0307077}}.

\bibitem{Brustein:1998kq}
R.~Brustein, M.~Gasperini, and G.~Veneziano, ``{Duality in cosmological
  perturbation theory},''
  \href{http://dx.doi.org/10.1016/S0370-2693(98)00576-0}{{\em Phys. Lett. B}
  {\bfseries 431} (1998) 277--285},
  \href{http://arxiv.org/abs/hep-th/9803018}{{\ttfamily arXiv:hep-th/9803018}}.

\bibitem{Starobinsky:1980te}
A.~A. Starobinsky, ``{A New Type of Isotropic Cosmological Models Without
  Singularity},'' \href{http://dx.doi.org/10.1016/0370-2693(80)90670-X}{{\em
  Phys. Lett. B} {\bfseries 91} (1980) 99--102}.

\bibitem{Mukhanov:1991zn}
V.~F. Mukhanov and R.~H. Brandenberger, ``{A Nonsingular universe},''
  \href{http://dx.doi.org/10.1103/PhysRevLett.68.1969}{{\em Phys. Rev. Lett.}
  {\bfseries 68} (1992) 1969--1972}.

\bibitem{Brandenberger:1993ef}
R.~H. Brandenberger, V.~F. Mukhanov, and A.~Sornborger, ``{A Cosmological
  theory without singularities},''
  \href{http://dx.doi.org/10.1103/PhysRevD.48.1629}{{\em Phys. Rev. D}
  {\bfseries 48} (1993) 1629--1642},
  \href{http://arxiv.org/abs/gr-qc/9303001}{{\ttfamily arXiv:gr-qc/9303001}}.

\bibitem{Novello:2008ra}
M.~Novello and S.~E.~P. Bergliaffa, ``{Bouncing Cosmologies},''
  \href{http://dx.doi.org/10.1016/j.physrep.2008.04.006}{{\em Phys. Rept.}
  {\bfseries 463} (2008) 127--213},
  \href{http://arxiv.org/abs/0802.1634}{{\ttfamily arXiv:0802.1634
  [astro-ph]}}.

\bibitem{Lilley:2015ksa}
M.~Lilley and P.~Peter, ``{Bouncing alternatives to inflation},''
  \href{http://dx.doi.org/10.1016/j.crhy.2015.08.009}{{\em Comptes Rendus
  Physique} {\bfseries 16} (2015) 1038--1047},
  \href{http://arxiv.org/abs/1503.06578}{{\ttfamily arXiv:1503.06578
  [astro-ph.CO]}}.

\bibitem{Battefeld:2014uga}
D.~Battefeld and P.~Peter, ``{A Critical Review of Classical Bouncing
  Cosmologies},'' \href{http://dx.doi.org/10.1016/j.physrep.2014.12.004}{{\em
  Phys. Rept.} {\bfseries 571} (2015) 1--66},
  \href{http://arxiv.org/abs/1406.2790}{{\ttfamily arXiv:1406.2790
  [astro-ph.CO]}}.

\bibitem{Peter:2008qz}
P.~Peter and N.~Pinto-Neto, ``{Cosmology without inflation},''
  \href{http://dx.doi.org/10.1103/PhysRevD.78.063506}{{\em Phys. Rev. D}
  {\bfseries 78} (2008) 063506},
  \href{http://arxiv.org/abs/0809.2022}{{\ttfamily arXiv:0809.2022 [gr-qc]}}.

\bibitem{Biswas:2005qr}
T.~Biswas, A.~Mazumdar, and W.~Siegel, ``{Bouncing universes in string-inspired
  gravity},'' \href{http://dx.doi.org/10.1088/1475-7516/2006/03/009}{{\em JCAP}
  {\bfseries 03} (2006) 009},
  \href{http://arxiv.org/abs/hep-th/0508194}{{\ttfamily arXiv:hep-th/0508194}}.

\bibitem{Bamba:2013fha}
K.~Bamba, A.~N. Makarenko, A.~N. Myagky, S.~Nojiri, and S.~D. Odintsov,
  ``{Bounce cosmology from $F(R)$ gravity and $F(R)$ bigravity},''
  \href{http://dx.doi.org/10.1088/1475-7516/2014/01/008}{{\em JCAP} {\bfseries
  01} (2014) 008}, \href{http://arxiv.org/abs/1309.3748}{{\ttfamily
  arXiv:1309.3748 [hep-th]}}.

\bibitem{Nojiri:2014zqa}
S.~Nojiri and S.~D. Odintsov, ``{Mimetic $F(R)$ gravity: inflation, dark energy
  and bounce},'' \href{http://arxiv.org/abs/1408.3561}{{\ttfamily
  arXiv:1408.3561 [hep-th]}}. [Erratum: Mod.Phys.Lett.A 29, 1450211 (2014)].

\bibitem{Bajardi:2020fxh}
F.~Bajardi, D.~Vernieri, and S.~Capozziello, ``{Bouncing Cosmology in f(Q)
  Symmetric Teleparallel Gravity},''
  \href{http://dx.doi.org/10.1140/epjp/s13360-020-00918-3}{{\em Eur. Phys. J.
  Plus} {\bfseries 135} no.~11, (2020) 912},
  \href{http://arxiv.org/abs/2011.01248}{{\ttfamily arXiv:2011.01248 [gr-qc]}}.

\bibitem{Bhargava:2020fhl}
P.~Bhargava, S.~Choudhury, S.~Chowdhury, A.~Mishara, S.~P. Selvam, S.~Panda,
  and G.~D. Pasquino, ``{Quantum aspects of chaos and complexity from bouncing
  cosmology: A study with two-mode single field squeezed state formalism},''
  \href{http://dx.doi.org/10.21468/SciPostPhysCore.4.4.026}{{\em SciPost Phys.
  Core} {\bfseries 4} (2021) 026},
  \href{http://arxiv.org/abs/2009.03893}{{\ttfamily arXiv:2009.03893
  [hep-th]}}.

\bibitem{Cai:2009in}
Y.-F. Cai and E.~N. Saridakis, ``{Non-singular cosmology in a model of
  non-relativistic gravity},''
  \href{http://dx.doi.org/10.1088/1475-7516/2009/10/020}{{\em JCAP} {\bfseries
  10} (2009) 020}, \href{http://arxiv.org/abs/0906.1789}{{\ttfamily
  arXiv:0906.1789 [hep-th]}}.

\bibitem{Cai:2012ag}
Y.-F. Cai, C.~Gao, and E.~N. Saridakis, ``{Bounce and cyclic cosmology in
  extended nonlinear massive gravity},''
  \href{http://dx.doi.org/10.1088/1475-7516/2012/10/048}{{\em JCAP} {\bfseries
  10} (2012) 048}, \href{http://arxiv.org/abs/1207.3786}{{\ttfamily
  arXiv:1207.3786 [astro-ph.CO]}}.

\bibitem{Shtanov:2002mb}
Y.~Shtanov and V.~Sahni, ``{Bouncing brane worlds},''
  \href{http://dx.doi.org/10.1016/S0370-2693(03)00179-5}{{\em Phys. Lett. B}
  {\bfseries 557} (2003) 1--6},
  \href{http://arxiv.org/abs/gr-qc/0208047}{{\ttfamily arXiv:gr-qc/0208047}}.

\bibitem{Ilyas:2020qja}
A.~Ilyas, M.~Zhu, Y.~Zheng, Y.-F. Cai, and E.~N. Saridakis, ``{DHOST Bounce},''
  \href{http://dx.doi.org/10.1088/1475-7516/2020/09/002}{{\em JCAP} {\bfseries
  09} (2020) 002}, \href{http://arxiv.org/abs/2002.08269}{{\ttfamily
  arXiv:2002.08269 [gr-qc]}}.

\bibitem{Ilyas:2020zcb}
A.~Ilyas, M.~Zhu, Y.~Zheng, and Y.-F. Cai, ``{Emergent Universe and Genesis
  from the DHOST Cosmology},''
  \href{http://dx.doi.org/10.1007/JHEP01(2021)141}{{\em JHEP} {\bfseries 01}
  (2021) 141}, \href{http://arxiv.org/abs/2009.10351}{{\ttfamily
  arXiv:2009.10351 [gr-qc]}}.

\bibitem{Zhu:2021whu}
M.~Zhu, A.~Ilyas, Y.~Zheng, Y.-F. Cai, and E.~N. Saridakis, ``{Scalar and
  tensor perturbations in DHOST bounce cosmology},''
  \href{http://dx.doi.org/10.1088/1475-7516/2021/11/045}{{\em JCAP} {\bfseries
  11} no.~11, (2021) 045}, \href{http://arxiv.org/abs/2108.01339}{{\ttfamily
  arXiv:2108.01339 [gr-qc]}}.

\bibitem{Banerjee:2016hom}
S.~Banerjee and E.~N. Saridakis, ``{Bounce and cyclic cosmology in weakly
  broken galileon theories},''
  \href{http://dx.doi.org/10.1103/PhysRevD.95.063523}{{\em Phys. Rev. D}
  {\bfseries 95} no.~6, (2017) 063523},
  \href{http://arxiv.org/abs/1604.06932}{{\ttfamily arXiv:1604.06932 [gr-qc]}}.

\bibitem{Saridakis:2018fth}
E.~N. Saridakis, S.~Banerjee, and R.~Myrzakulov, ``{Bounce and cyclic cosmology
  in new gravitational scalar-tensor theories},''
  \href{http://dx.doi.org/10.1103/PhysRevD.98.063513}{{\em Phys. Rev. D}
  {\bfseries 98} no.~6, (2018) 063513},
  \href{http://arxiv.org/abs/1807.00346}{{\ttfamily arXiv:1807.00346 [gr-qc]}}.

\bibitem{Barca:2021qdn}
G.~Barca, E.~Giovannetti, and G.~Montani, ``{An Overview on the Nature of the
  Bounce in LQC and PQM},''
  \href{http://dx.doi.org/10.3390/universe7090327}{{\em Universe} {\bfseries 7}
  no.~9, (2021) 327}, \href{http://arxiv.org/abs/2109.08645}{{\ttfamily
  arXiv:2109.08645 [gr-qc]}}.

\bibitem{Wilson-Ewing:2012lmx}
E.~Wilson-Ewing, ``{The Matter Bounce Scenario in Loop Quantum Cosmology},''
  \href{http://dx.doi.org/10.1088/1475-7516/2013/03/026}{{\em JCAP} {\bfseries
  03} (2013) 026}, \href{http://arxiv.org/abs/1211.6269}{{\ttfamily
  arXiv:1211.6269 [gr-qc]}}.

\bibitem{K:2023gsi}
R.~K and V.~Sreenath, ``{Estimation of imprints of the bounce in loop quantum
  cosmology on the bispectra of cosmic microwave background},''
  \href{http://dx.doi.org/10.1088/1475-7516/2023/08/014}{{\em JCAP} {\bfseries
  08} (2023) 014}, \href{http://arxiv.org/abs/2301.05406}{{\ttfamily
  arXiv:2301.05406 [astro-ph.CO]}}.

\bibitem{Agullo:2020cvg}
I.~Agullo, D.~Kranas, and V.~Sreenath, ``{Large scale anomalies in the CMB and
  non-Gaussianity in bouncing cosmologies},''
  \href{http://dx.doi.org/10.1088/1361-6382/abc521}{{\em Class. Quant. Grav.}
  {\bfseries 38} no.~6, (2021) 065010},
  \href{http://arxiv.org/abs/2006.09605}{{\ttfamily arXiv:2006.09605
  [astro-ph.CO]}}.

\bibitem{Agullo:2020fbw}
I.~Agullo, D.~Kranas, and V.~Sreenath, ``{Anomalies in the CMB from a cosmic
  bounce},'' \href{http://dx.doi.org/10.1007/s10714-020-02778-9}{{\em Gen. Rel.
  Grav.} {\bfseries 53} no.~2, (2021) 17},
  \href{http://arxiv.org/abs/2005.01796}{{\ttfamily arXiv:2005.01796
  [astro-ph.CO]}}.

\bibitem{Agullo:2020wur}
I.~Agullo, J.~Olmedo, and V.~Sreenath, ``{Predictions for the Cosmic Microwave
  Background from an Anisotropic Quantum Bounce},''
  \href{http://dx.doi.org/10.1103/PhysRevLett.124.251301}{{\em Phys. Rev.
  Lett.} {\bfseries 124} no.~25, (2020) 251301},
  \href{http://arxiv.org/abs/2003.02304}{{\ttfamily arXiv:2003.02304 [gr-qc]}}.

\bibitem{Stargen:2016cft}
D.~J. Stargen, V.~Sreenath, and L.~Sriramkumar, ``{Quantum-to-classical
  transition and imprints of continuous spontaneous localization in classical
  bouncing universes},''
  \href{http://dx.doi.org/10.1142/S0218271821500498}{{\em Int. J. Mod. Phys. D}
  {\bfseries 30} no.~07, (2021) 2150049},
  \href{http://arxiv.org/abs/1605.07311}{{\ttfamily arXiv:1605.07311 [gr-qc]}}.

\bibitem{Sriramkumar:2015yza}
L.~Sriramkumar, K.~Atmjeet, and R.~K. Jain, ``{Generation of scale invariant
  magnetic fields in bouncing universes},''
  \href{http://dx.doi.org/10.1088/1475-7516/2015/09/010}{{\em JCAP} {\bfseries
  09} (2015) 010}, \href{http://arxiv.org/abs/1504.06853}{{\ttfamily
  arXiv:1504.06853 [astro-ph.CO]}}.

\bibitem{Banerjee:2022gpy}
I.~Banerjee, T.~Paul, and S.~SenGupta, ``{Aspects of non-singular bounce in
  modified gravity theories},''
  \href{http://dx.doi.org/10.1007/s10714-022-02988-3}{{\em Gen. Rel. Grav.}
  {\bfseries 54} no.~10, (2022) 119},
  \href{http://arxiv.org/abs/2205.05283}{{\ttfamily arXiv:2205.05283 [gr-qc]}}.

\bibitem{Paul:2022mup}
T.~Paul and S.~SenGupta, ``{Ekpyrotic bounce driven by Kalb\textendash{}Ramond
  field},'' \href{http://dx.doi.org/10.1016/j.dark.2023.101236}{{\em Phys. Dark
  Univ.} {\bfseries 41} (2023) 101236},
  \href{http://arxiv.org/abs/2202.13186}{{\ttfamily arXiv:2202.13186 [gr-qc]}}.

\bibitem{Odintsov:2021yva}
S.~D. Odintsov, T.~Paul, I.~Banerjee, R.~Myrzakulov, and S.~SenGupta,
  ``{Unifying an asymmetric bounce to the dark energy in
  Chern\textendash{}Simons F(R) gravity},''
  \href{http://dx.doi.org/10.1016/j.dark.2021.100864}{{\em Phys. Dark Univ.}
  {\bfseries 33} (2021) 100864},
  \href{http://arxiv.org/abs/2109.00345}{{\ttfamily arXiv:2109.00345 [gr-qc]}}.

\bibitem{Banerjee:2020uil}
I.~Banerjee, T.~Paul, and S.~SenGupta, ``{Bouncing cosmology in a curved
  braneworld},'' \href{http://dx.doi.org/10.1088/1475-7516/2021/02/041}{{\em
  JCAP} {\bfseries 02} (2021) 041},
  \href{http://arxiv.org/abs/2011.11886}{{\ttfamily arXiv:2011.11886 [gr-qc]}}.

\bibitem{Das:2017jrl}
A.~Das, D.~Maity, T.~Paul, and S.~SenGupta, ``{Bouncing cosmology from warped
  extra dimensional scenario},''
  \href{http://dx.doi.org/10.1140/epjc/s10052-017-5396-2}{{\em Eur. Phys. J. C}
  {\bfseries 77} no.~12, (2017) 813},
  \href{http://arxiv.org/abs/1706.00950}{{\ttfamily arXiv:1706.00950
  [hep-th]}}.

\bibitem{Pan:2024ydt}
S.~Pan, Y.~Cai, and Y.-S. Piao, ``{Climbing over the potential barrier during
  inflation via null energy condition violation},''
  \href{http://arxiv.org/abs/2404.12655}{{\ttfamily arXiv:2404.12655
  [astro-ph.CO]}}.

\bibitem{Colas:2024xjy}
T.~Colas, C.~de~Rham, and G.~Kaplanek, ``{Decoherence out of fire: purity loss
  in expanding and contracting universes},''
  \href{http://dx.doi.org/10.1088/1475-7516/2024/05/025}{{\em JCAP} {\bfseries
  05} (2024) 025}, \href{http://arxiv.org/abs/2401.02832}{{\ttfamily
  arXiv:2401.02832 [hep-th]}}.

\bibitem{Piao:2003zm}
Y.-S. Piao, B.~Feng, and X.-m. Zhang, ``{Suppressing CMB quadrupole with a
  bounce from contracting phase to inflation},''
  \href{http://dx.doi.org/10.1103/PhysRevD.69.103520}{{\em Phys. Rev. D}
  {\bfseries 69} (2004) 103520},
  \href{http://arxiv.org/abs/hep-th/0310206}{{\ttfamily arXiv:hep-th/0310206}}.

\bibitem{Cai:2017dyi}
Y.~Cai and Y.-S. Piao, ``{A covariant Lagrangian for stable nonsingular
  bounce},'' \href{http://dx.doi.org/10.1007/JHEP09(2017)027}{{\em JHEP}
  {\bfseries 09} (2017) 027}, \href{http://arxiv.org/abs/1705.03401}{{\ttfamily
  arXiv:1705.03401 [gr-qc]}}.

\bibitem{Cai:2017pga}
Y.~Cai, Y.-T. Wang, J.-Y. Zhao, and Y.-S. Piao, ``{Primordial perturbations
  with pre-inflationary bounce},''
  \href{http://dx.doi.org/10.1103/PhysRevD.97.103535}{{\em Phys. Rev. D}
  {\bfseries 97} no.~10, (2018) 103535},
  \href{http://arxiv.org/abs/1709.07464}{{\ttfamily arXiv:1709.07464
  [astro-ph.CO]}}.

\bibitem{Cai:2015nya}
Y.~Cai, Y.-T. Wang, and Y.-S. Piao, ``{Preinflationary primordial
  perturbations},'' \href{http://dx.doi.org/10.1103/PhysRevD.92.023518}{{\em
  Phys. Rev. D} {\bfseries 92} no.~2, (2015) 023518},
  \href{http://arxiv.org/abs/1501.01730}{{\ttfamily arXiv:1501.01730
  [astro-ph.CO]}}.

\bibitem{Cai:2019hge}
Y.~Cai and Y.-S. Piao, ``{Pre-inflation and trans-Planckian censorship},''
  \href{http://dx.doi.org/10.1007/s11433-020-1573-5}{{\em Sci. China Phys.
  Mech. Astron.} {\bfseries 63} no.~11, (2020) 110411},
  \href{http://arxiv.org/abs/1909.12719}{{\ttfamily arXiv:1909.12719 [gr-qc]}}.

\bibitem{Zhu:2023lbf}
M.~Zhu, G.~Ye, and Y.~Cai, ``{Pulsar timing array observations as possible
  hints for nonsingular cosmology},''
  \href{http://dx.doi.org/10.1140/epjc/s10052-023-11963-4}{{\em Eur. Phys. J.
  C} {\bfseries 83} no.~9, (2023) 816},
  \href{http://arxiv.org/abs/2307.16211}{{\ttfamily arXiv:2307.16211
  [astro-ph.CO]}}.

\bibitem{Weinberg:2008hq}
S.~Weinberg, ``{Effective Field Theory for Inflation},''
  \href{http://dx.doi.org/10.1103/PhysRevD.77.123541}{{\em Phys. Rev. D}
  {\bfseries 77} (2008) 123541},
  \href{http://arxiv.org/abs/0804.4291}{{\ttfamily arXiv:0804.4291 [hep-th]}}.

\bibitem{Cheung:2007st}
C.~Cheung, P.~Creminelli, A.~L. Fitzpatrick, J.~Kaplan, and L.~Senatore, ``{The
  Effective Field Theory of Inflation},''
  \href{http://dx.doi.org/10.1088/1126-6708/2008/03/014}{{\em JHEP} {\bfseries
  03} (2008) 014}, \href{http://arxiv.org/abs/0709.0293}{{\ttfamily
  arXiv:0709.0293 [hep-th]}}.

\bibitem{Choudhury:2017glj}
S.~Choudhury, ``{CMB from EFT},''
  \href{http://dx.doi.org/10.3390/universe5060155}{{\em Universe} {\bfseries 5}
  no.~6, (2019) 155}, \href{http://arxiv.org/abs/1712.04766}{{\ttfamily
  arXiv:1712.04766 [hep-th]}}.

\bibitem{Delacretaz:2016nhw}
L.~V. Delacretaz, V.~Gorbenko, and L.~Senatore, ``{The Supersymmetric Effective
  Field Theory of Inflation},''
  \href{http://dx.doi.org/10.1007/JHEP03(2017)063}{{\em JHEP} {\bfseries 03}
  (2017) 063}, \href{http://arxiv.org/abs/1610.04227}{{\ttfamily
  arXiv:1610.04227 [hep-th]}}.

\bibitem{Chen:2016nrs}
X.~Chen, Y.~Wang, and Z.-Z. Xianyu, ``{Loop Corrections to Standard Model
  Fields in Inflation},'' \href{http://dx.doi.org/10.1007/JHEP08(2016)051}{{\em
  JHEP} {\bfseries 08} (2016) 051},
  \href{http://arxiv.org/abs/1604.07841}{{\ttfamily arXiv:1604.07841
  [hep-th]}}.

\bibitem{Baumann:2019ghk}
D.~Baumann, D.~Green, and T.~Hartman, ``{Dynamical Constraints on RG Flows and
  Cosmology},'' \href{http://dx.doi.org/10.1007/JHEP12(2019)134}{{\em JHEP}
  {\bfseries 12} (2019) 134}, \href{http://arxiv.org/abs/1906.10226}{{\ttfamily
  arXiv:1906.10226 [hep-th]}}.

\bibitem{Boyanovsky:1998aa}
D.~Boyanovsky, H.~J. de~Vega, R.~Holman, and M.~Simionato, ``{Dynamical
  renormalization group resummation of finite temperature infrared
  divergences},'' \href{http://dx.doi.org/10.1103/PhysRevD.60.065003}{{\em
  Phys. Rev. D} {\bfseries 60} (1999) 065003},
  \href{http://arxiv.org/abs/hep-ph/9809346}{{\ttfamily arXiv:hep-ph/9809346}}.

\bibitem{Boyanovsky:2001ty}
D.~Boyanovsky, H.~J. De~Vega, D.~S. Lee, S.-Y. Wang, and H.~L. Yu, ``{Dynamical
  renormalization group approach to the Altarelli-Parisi equations},''
  \href{http://dx.doi.org/10.1103/PhysRevD.65.045014}{{\em Phys. Rev. D}
  {\bfseries 65} (2002) 045014},
  \href{http://arxiv.org/abs/hep-ph/0108180}{{\ttfamily arXiv:hep-ph/0108180}}.

\bibitem{Boyanovsky:2003ui}
D.~Boyanovsky and H.~J. de~Vega, ``{Dynamical renormalization group approach to
  relaxation in quantum field theory},''
  \href{http://dx.doi.org/10.1016/S0003-4916(03)00115-5}{{\em Annals Phys.}
  {\bfseries 307} (2003) 335--371},
  \href{http://arxiv.org/abs/hep-ph/0302055}{{\ttfamily arXiv:hep-ph/0302055}}.

\bibitem{Burgess:2015ajz}
C.~P. Burgess, R.~Holman, and G.~Tasinato, ``{Open EFTs, IR effects
  \textbackslash{}\& late-time resummations: systematic corrections in
  stochastic inflation},''
  \href{http://dx.doi.org/10.1007/JHEP01(2016)153}{{\em JHEP} {\bfseries 01}
  (2016) 153}, \href{http://arxiv.org/abs/1512.00169}{{\ttfamily
  arXiv:1512.00169 [gr-qc]}}.

\bibitem{Burgess:2014eoa}
C.~P. Burgess, R.~Holman, G.~Tasinato, and M.~Williams, ``{EFT Beyond the
  Horizon: Stochastic Inflation and How Primordial Quantum Fluctuations Go
  Classical},'' \href{http://dx.doi.org/10.1007/JHEP03(2015)090}{{\em JHEP}
  {\bfseries 03} (2015) 090}, \href{http://arxiv.org/abs/1408.5002}{{\ttfamily
  arXiv:1408.5002 [hep-th]}}.

\bibitem{Burgess:2009bs}
C.~P. Burgess, L.~Leblond, R.~Holman, and S.~Shandera, ``{Super-Hubble de
  Sitter Fluctuations and the Dynamical RG},''
  \href{http://dx.doi.org/10.1088/1475-7516/2010/03/033}{{\em JCAP} {\bfseries
  03} (2010) 033}, \href{http://arxiv.org/abs/0912.1608}{{\ttfamily
  arXiv:0912.1608 [hep-th]}}.

\bibitem{Dias:2012qy}
M.~Dias, R.~H. Ribeiro, and D.~Seery, ``{The \ensuremath{\delta}N formula is
  the dynamical renormalization group},''
  \href{http://dx.doi.org/10.1088/1475-7516/2013/10/062}{{\em JCAP} {\bfseries
  10} (2013) 062}, \href{http://arxiv.org/abs/1210.7800}{{\ttfamily
  arXiv:1210.7800 [astro-ph.CO]}}.

\bibitem{Chaykov:2022zro}
S.~Chaykov, N.~Agarwal, S.~Bahrami, and R.~Holman, ``{Loop corrections in
  Minkowski spacetime away from equilibrium. Part I. Late-time resummations},''
  \href{http://dx.doi.org/10.1007/JHEP02(2023)093}{{\em JHEP} {\bfseries 02}
  (2023) 093}, \href{http://arxiv.org/abs/2206.11288}{{\ttfamily
  arXiv:2206.11288 [hep-th]}}.

\bibitem{Chaykov:2022pwd}
S.~Chaykov, N.~Agarwal, S.~Bahrami, and R.~Holman, ``{Loop corrections in
  Minkowski spacetime away from equilibrium. Part II. Finite-time results},''
  \href{http://dx.doi.org/10.1007/JHEP02(2023)094}{{\em JHEP} {\bfseries 02}
  (2023) 094}, \href{http://arxiv.org/abs/2206.11289}{{\ttfamily
  arXiv:2206.11289 [hep-th]}}.

\bibitem{Planck:2018jri}
{\bfseries Planck} Collaboration, Y.~Akrami {\em et~al.}, ``{Planck 2018
  results. X. Constraints on inflation},''
  \href{http://dx.doi.org/10.1051/0004-6361/201833887}{{\em Astron. Astrophys.}
  {\bfseries 641} (2020) A10},
  \href{http://arxiv.org/abs/1807.06211}{{\ttfamily arXiv:1807.06211
  [astro-ph.CO]}}.

\bibitem{Cyr:2023pgw}
B.~Cyr, T.~Kite, J.~Chluba, J.~C. Hill, D.~Jeong, S.~K. Acharya, B.~Bolliet,
  and S.~P. Patil, ``{Disentangling the primordial nature of stochastic
  gravitational wave backgrounds with CMB spectral distortions},''
  \href{http://dx.doi.org/10.1093/mnras/stad3861}{{\em Mon. Not. Roy. Astron.
  Soc.} {\bfseries 528} no.~1, (2024) 883--897},
  \href{http://arxiv.org/abs/2309.02366}{{\ttfamily arXiv:2309.02366
  [astro-ph.CO]}}.

\bibitem{EPTA:2023xxk}
{\bfseries EPTA} Collaboration, J.~Antoniadis {\em et~al.}, ``{The second data
  release from the European Pulsar Timing Array: V. Implications for massive
  black holes, dark matter and the early Universe},''
  \href{http://arxiv.org/abs/2306.16227}{{\ttfamily arXiv:2306.16227
  [astro-ph.CO]}}.

\bibitem{bird2011minimally}
S.~Bird, H.~V. Peiris, M.~Viel, and L.~Verde, ``Minimally parametric power
  spectrum reconstruction from the lyman $\alpha$ forest,'' {\em Monthly
  Notices of the Royal Astronomical Society} {\bfseries 413} no.~3, (2011)
  1717--1728.

\bibitem{Jeong:2014gna}
D.~Jeong, J.~Pradler, J.~Chluba, and M.~Kamionkowski, ``{Silk damping at a
  redshift of a billion: a new limit on small-scale adiabatic perturbations},''
  \href{http://dx.doi.org/10.1103/PhysRevLett.113.061301}{{\em Phys. Rev.
  Lett.} {\bfseries 113} (2014) 061301},
  \href{http://arxiv.org/abs/1403.3697}{{\ttfamily arXiv:1403.3697
  [astro-ph.CO]}}.

\bibitem{Mahbub:2019uhl}
R.~Mahbub, ``{Primordial black hole formation in inflationary
  $\alpha$-attractor models},''
  \href{http://dx.doi.org/10.1103/PhysRevD.101.023533}{{\em Phys. Rev. D}
  {\bfseries 101} no.~2, (2020) 023533},
  \href{http://arxiv.org/abs/1910.10602}{{\ttfamily arXiv:1910.10602
  [astro-ph.CO]}}.

\bibitem{Harada:2013epa}
T.~Harada, C.-M. Yoo, and K.~Kohri, ``{Threshold of primordial black hole
  formation},'' \href{http://dx.doi.org/10.1103/PhysRevD.88.084051}{{\em Phys.
  Rev. D} {\bfseries 88} no.~8, (2013) 084051},
  \href{http://arxiv.org/abs/1309.4201}{{\ttfamily arXiv:1309.4201
  [astro-ph.CO]}}. [Erratum: Phys.Rev.D 89, 029903 (2014)].

\bibitem{Shibata:1999zs}
M.~Shibata and M.~Sasaki, ``{Black hole formation in the Friedmann universe:
  Formulation and computation in numerical relativity},''
  \href{http://dx.doi.org/10.1103/PhysRevD.60.084002}{{\em Phys. Rev. D}
  {\bfseries 60} (1999) 084002},
  \href{http://arxiv.org/abs/gr-qc/9905064}{{\ttfamily arXiv:gr-qc/9905064}}.

\bibitem{Musco:2008hv}
I.~Musco, J.~C. Miller, and A.~G. Polnarev, ``{Primordial black hole formation
  in the radiative era: Investigation of the critical nature of the
  collapse},'' \href{http://dx.doi.org/10.1088/0264-9381/26/23/235001}{{\em
  Class. Quant. Grav.} {\bfseries 26} (2009) 235001},
  \href{http://arxiv.org/abs/0811.1452}{{\ttfamily arXiv:0811.1452 [gr-qc]}}.

\bibitem{Polnarev:2006aa}
A.~G. Polnarev and I.~Musco, ``{Curvature profiles as initial conditions for
  primordial black hole formation},''
  \href{http://dx.doi.org/10.1088/0264-9381/24/6/003}{{\em Class. Quant. Grav.}
  {\bfseries 24} (2007) 1405--1432},
  \href{http://arxiv.org/abs/gr-qc/0605122}{{\ttfamily arXiv:gr-qc/0605122}}.

\bibitem{Escriva:2019phb}
A.~Escriv{\`a}, C.~Germani, and R.~K. Sheth, ``{Universal threshold for
  primordial black hole formation},''
  \href{http://dx.doi.org/10.1103/PhysRevD.101.044022}{{\em Phys. Rev. D}
  {\bfseries 101} no.~4, (2020) 044022},
  \href{http://arxiv.org/abs/1907.13311}{{\ttfamily arXiv:1907.13311 [gr-qc]}}.

\bibitem{Motohashi:2017kbs}
H.~Motohashi and W.~Hu, ``{Primordial Black Holes and Slow-Roll Violation},''
  \href{http://dx.doi.org/10.1103/PhysRevD.96.063503}{{\em Phys. Rev. D}
  {\bfseries 96} no.~6, (2017) 063503},
  \href{http://arxiv.org/abs/1706.06784}{{\ttfamily arXiv:1706.06784
  [astro-ph.CO]}}.

\bibitem{Ballesteros:2017fsr}
G.~Ballesteros and M.~Taoso, ``{Primordial black hole dark matter from single
  field inflation},'' \href{http://dx.doi.org/10.1103/PhysRevD.97.023501}{{\em
  Phys. Rev. D} {\bfseries 97} no.~2, (2018) 023501},
  \href{http://arxiv.org/abs/1709.05565}{{\ttfamily arXiv:1709.05565
  [hep-ph]}}.

\bibitem{Young:2019osy}
S.~Young, ``{The primordial black hole formation criterion re-examined:
  Parametrisation, timing and the choice of window function},''
  \href{http://dx.doi.org/10.1142/S0218271820300025}{{\em Int. J. Mod. Phys. D}
  {\bfseries 29} no.~02, (2019) 2030002},
  \href{http://arxiv.org/abs/1905.01230}{{\ttfamily arXiv:1905.01230
  [astro-ph.CO]}}.

\bibitem{Ando:2018qdb}
K.~Ando, K.~Inomata, and M.~Kawasaki, ``{Primordial black holes and
  uncertainties in the choice of the window function},''
  \href{http://dx.doi.org/10.1103/PhysRevD.97.103528}{{\em Phys. Rev. D}
  {\bfseries 97} no.~10, (2018) 103528},
  \href{http://arxiv.org/abs/1802.06393}{{\ttfamily arXiv:1802.06393
  [astro-ph.CO]}}.

\bibitem{Young:2019yug}
S.~Young, I.~Musco, and C.~T. Byrnes, ``{Primordial black hole formation and
  abundance: contribution from the non-linear relation between the density and
  curvature perturbation},''
  \href{http://dx.doi.org/10.1088/1475-7516/2019/11/012}{{\em JCAP} {\bfseries
  11} (2019) 012}, \href{http://arxiv.org/abs/1904.00984}{{\ttfamily
  arXiv:1904.00984 [astro-ph.CO]}}.

\bibitem{DeLuca:2019qsy}
V.~De~Luca, G.~Franciolini, A.~Kehagias, M.~Peloso, A.~Riotto, and C.~{\"U}nal,
  ``{The Ineludible non-Gaussianity of the Primordial Black Hole Abundance},''
  \href{http://dx.doi.org/10.1088/1475-7516/2019/07/048}{{\em JCAP} {\bfseries
  07} (2019) 048}, \href{http://arxiv.org/abs/1904.00970}{{\ttfamily
  arXiv:1904.00970 [astro-ph.CO]}}.

\bibitem{Musco:2018rwt}
I.~Musco, ``{Threshold for primordial black holes: Dependence on the shape of
  the cosmological perturbations},''
  \href{http://dx.doi.org/10.1103/PhysRevD.100.123524}{{\em Phys. Rev. D}
  {\bfseries 100} no.~12, (2019) 123524},
  \href{http://arxiv.org/abs/1809.02127}{{\ttfamily arXiv:1809.02127 [gr-qc]}}.

\bibitem{Musco:2020jjb}
I.~Musco, V.~De~Luca, G.~Franciolini, and A.~Riotto, ``{Threshold for
  primordial black holes. II. A simple analytic prescription},''
  \href{http://dx.doi.org/10.1103/PhysRevD.103.063538}{{\em Phys. Rev. D}
  {\bfseries 103} no.~6, (2021) 063538},
  \href{http://arxiv.org/abs/2011.03014}{{\ttfamily arXiv:2011.03014
  [astro-ph.CO]}}.

\bibitem{Cole:2023wyx}
P.~S. Cole, A.~D. Gow, C.~T. Byrnes, and S.~P. Patil, ``{Primordial black holes
  from single-field inflation: a fine-tuning audit},''
  \href{http://dx.doi.org/10.1088/1475-7516/2023/08/031}{{\em JCAP} {\bfseries
  08} (2023) 031}, \href{http://arxiv.org/abs/2304.01997}{{\ttfamily
  arXiv:2304.01997 [astro-ph.CO]}}.

\bibitem{Profumo:2026qpn}
S.~Profumo, ``{Are Primordial Black Holes a Natural Dark Matter Candidate?},''
  \href{http://arxiv.org/abs/2606.12775}{{\ttfamily arXiv:2606.12775
  [hep-ph]}}.

\bibitem{Profumo:2026kfy}
S.~Profumo, ``{What Naturalness Measures: Fine-Tuning and Informational
  Invariants in Cosmology and Dark Matter},''
  \href{http://arxiv.org/abs/2606.29660}{{\ttfamily arXiv:2606.29660
  [physics.hist-ph]}}.

\bibitem{Carr:2011hv}
B.~J. Carr and A.~A. Coley, ``{Persistence of black holes through a
  cosmological bounce},''
  \href{http://dx.doi.org/10.1142/S0218271811020640}{{\em Int. J. Mod. Phys. D}
  {\bfseries 20} (2011) 2733--2738},
  \href{http://arxiv.org/abs/1104.3796}{{\ttfamily arXiv:1104.3796
  [astro-ph.CO]}}.

\bibitem{Quintin:2016qro}
J.~Quintin and R.~H. Brandenberger, ``{Black hole formation in a contracting
  universe},'' \href{http://dx.doi.org/10.1088/1475-7516/2016/11/029}{{\em
  JCAP} {\bfseries 11} (2016) 029},
  \href{http://arxiv.org/abs/1609.02556}{{\ttfamily arXiv:1609.02556
  [astro-ph.CO]}}.

\bibitem{Chen:2016kjx}
J.-W. Chen, J.~Liu, H.-L. Xu, and Y.-F. Cai, ``{Tracing Primordial Black Holes
  in Nonsingular Bouncing Cosmology},''
  \href{http://dx.doi.org/10.1016/j.physletb.2017.03.036}{{\em Phys. Lett. B}
  {\bfseries 769} (2017) 561--568},
  \href{http://arxiv.org/abs/1609.02571}{{\ttfamily arXiv:1609.02571 [gr-qc]}}.

\bibitem{Clifton:2017hvg}
T.~Clifton, B.~Carr, and A.~Coley, ``{Persistent Black Holes in Bouncing
  Cosmologies},'' \href{http://dx.doi.org/10.1088/1361-6382/aa6dbb}{{\em Class.
  Quant. Grav.} {\bfseries 34} no.~13, (2017) 135005},
  \href{http://arxiv.org/abs/1701.05750}{{\ttfamily arXiv:1701.05750 [gr-qc]}}.

\bibitem{Banerjee:2022xft}
S.~Banerjee, T.~Papanikolaou, and E.~N. Saridakis, ``{Constraining F(R)
  bouncing cosmologies through primordial black holes},''
  \href{http://dx.doi.org/10.1103/PhysRevD.106.124012}{{\em Phys. Rev. D}
  {\bfseries 106} no.~12, (2022) 124012},
  \href{http://arxiv.org/abs/2206.01150}{{\ttfamily arXiv:2206.01150 [gr-qc]}}.

\bibitem{Chen:2022usd}
J.-W. Chen, M.~Zhu, S.-F. Yan, Q.-Q. Wang, and Y.-F. Cai, ``{Enhance primordial
  black hole abundance through the non-linear processes around bounce point},''
  \href{http://dx.doi.org/10.1088/1475-7516/2023/01/015}{{\em JCAP} {\bfseries
  01} (2023) 015}, \href{http://arxiv.org/abs/2207.14532}{{\ttfamily
  arXiv:2207.14532 [astro-ph.CO]}}.

\bibitem{Papanikolaou:2024fzf}
T.~Papanikolaou, S.~Banerjee, Y.-F. Cai, S.~Capozziello, and E.~N. Saridakis,
  ``{Primordial black holes and induced gravitational waves in non-singular
  matter bouncing cosmology},''
  \href{http://arxiv.org/abs/2404.03779}{{\ttfamily arXiv:2404.03779 [gr-qc]}}.

\end{thebibliography}\endgroup
\bibliographystyle{utphys}


\end{document}